# A general formal memory framework in Coq for verifying the properties of programs based on higher-order logic theorem proving with increased automation, consistency, and reusability

Zheng Yang[1*]　　　Hang Lei

zyang.uestc@gmail.com　　　hlei@uestc.edu.cn

[1]School of Information and Software Engineering, University of Electronic Science and Technology of China,

No.4, Section 2, North Jianshe Road, 610054, Sichuan, Chengdu, P.R. China.

**Abstract** In recent years, a number of lightweight programs have been deployed in critical domains, such as in smart contracts based on blockchain technology. Therefore, the security and reliability of such programs should be guaranteed by the most credible technology. Higher-order logic theorem proving is one of the most reliable technologies for verifying the properties of programs. However, programs may be developed by different high level programming languages, and a general, extensible, and reusable formal memory (GERM) framework that can simultaneously support different formal verification specifications, particularly at the code level, is presently unavailable for verifying the properties of programs. Therefore, the present work proposes a GERM framework to fill this gap. The framework simulates physical memory hardware structure, including a low-level formal memory space, and provides a set of simple, nonintrusive application programming interfaces and assistant tools using Coq that can support different formal verification specifications simultaneously. The proposed GERM framework is independent and customizable, and was verified entirely in Coq. We also present an extension of Curry-Howard isomorphism, denoted as execution-verification isomorphism (EVI), which combines symbolic execution and theorem proving for increasing the degree of automation in higher-order logic theorem proving assistant tools. We also implement a toy intermediate programming language in a generalized algebraic datatypes style and a formal interpreter in Coq based on the GERM framework. These implementations are then employed to demonstrate the application of EVI to a simple code segment. This work is the first step in our project to build a general and powerful formal symbolic process virtual machine for certifying and verifying smart contracts operating on the blockchain platform easily and semi-automatically without consistency problems.

**Keywords**: formal framework, programming language, higher-order logic theorem proving, Coq.

## 1. Introduction

In recent years, a number of lightweight programs have been deployed in critical domains, such as in smart contracts based on blockchain technology [1]. Therefore, verifying the security and reliability of such programs in the most rigorous manner available is crucial. Higher-order logic theorem proving is one of the most rigorous technologies for verifying the properties of programs. This involves establishing a formal model of a software system, and then verifying the system according to a mathematical proof of the formal model. In a standard approach, researchers can abstract a specific formal model for target software systems manually with the help of proof assistants [3]. This type of formal verification technology has many advantages. For example, it provides sufficient freedom and flexibility for designing formal models using higher-order logic theories, and can abstract and express very complex systems. However, numerous problems are encountered when applying theorem proving technology to program verification. For example, the abstraction and translation processes are completed manually. As such, the formal models obtained are dependent on the experience, knowledge, and proficiency of researchers. This invariably leads to a general lack of consistency between the formalization results obtained by different researchers. This lack of consistency is exacerbated by the fact that at present, verifiers employed in program verification cannot find a standard general formal state model to define intermediate states, which could be used to derive invariant models. Thus verifiers usually choose different algorithms to define the intermediate states in different situations. Unfortunately, the consistence among different state model is difficult to be guaranteed, even though they are used to prove identical theorems. Furthermore, different algorithms need different methods to derive, which is one of the main reasons obstructing the reusability and automation of theorem proving. Moreover, due to the lack of a unified state model, if the formal model is based on axiomatic semantics, such as Hoare logic [4] or separation logic [5], the modification of a single logical statement may force a large adjustment in the definition of states, or even force the rebuilding the formal model manually. These factors severely limit the universality of a conventionally derived formal model, and thereby severely limit its reusability. One of the most troubling problems associated with of this type of formal verification technology is that the consistency between the formal model and the original program cannot be ensured formally. As a result, the formal model runs the risk of misunderstanding the source program logic and implementation, and may import vulnerabilities not existing in the original program, or remove vulnerabilities existing in the source code as an unintended result of the abstraction and translation processes. Finally, the formalization workload involved is very heavy. For example, the complete verification of the seL4 operating system (OS) kernel [6] required a total of 11 man-years, and the ratio of the original code of a complete, general-purpose concurrent OS kernel to the verification code of the CertiKOS [7] project was nearly 1:50. Although many higher-order theorem proving assistants provide a "*tactic*" mechanism [14] to help users design proving tactics to simplify programs evaluation process and construct proofs automatically, on amount of the differences among different formal models caused by above problems, it is hard to design tactics to verify formal models full-automatically.

One of the available solutions for addressing the above issues surrounding reusability, consistency, and automation is to design a formal symbolic process virtual machine (FSPVM) like KLEE [39], but developed in a higher-order theorem proving system, which can symbolically execute real world programs and verify their properties automatically using the execution result. However, if we want to implement it, we must overcome the following challenges.

The first challenge is developing an independent general formal memory model. It is the basis to construct a logic operating environment with the higher-order theorem proving system. It should be easily to support arbitrary high-level formal specifications to record their logic invariants and represent intermediate states during verification. It contributes to the reusability problem. Because it unifies the verification intermediate states and it can be used as the standard state model reused in different program verification models.

The second challenge is formalizing real world programming languages as an extensible intermediate programming language (IPL) and mechanizing IPL into the logic operating environment. The formal syntax and semantics of IPL should be equivalent with the respective real world target programming languages'. The IPL is for the reusability and consistency problems. Because, it standardizes the process of building a formal model for programs that the equivalent formal version programs written in IPL can be served as their formal models without abstracting or rebuilding.

The third challenge is developing a formal verified execution engine such as the formal interpreter based on the challenge 1 and 2. The execution engine should be able to automatically execute the formal version of programs written in IPL in the logic operating environment. And it is for the automation problem.

The fourth challenge is giving a theory for combining above symbolic execution elements and higher-order theorem proving to verify programs automatically, which contributes to reusability, consistency, and automation problems.

In this paper we have solved the challenge 1 and 4, and the present study makes the following contributions.

- We design a general, extensible, and reusable formal memory (GERM) framework based on higher-order logic using Coq. It includes a formal memory space, and provides a set of simple and nonintrusive memory management APIs and a set of assistant tools. The GERM framework can express the interaction relationships between special and normal memory blocks. One the one hand, the framework functions independently of higher level specifications, so it can be used to represent intermediate states of any high-level specifications designated by general users, which facilitates the reuse of intermediate representations in different high-level formal verification models. On the other hand, the framework can be used as an operating environment to facilitate automated higher-order logic theorem proving.
- We present a novel extension of Curry-Howard isomorphism (CHI), denoted herein as execution-verification isomorphism (EVI), which can combine theorem proving and symbolic execution technology in the operating environment of the GERM framework to facilitate automated higher-order logic theorem proving. The use of EVI makes it possible to execute a real world program logically while simultaneously verifying the properties of the program automatically in Coq or using another proof assistant that supports higher-order logic proving based on CHI without suffering inconsistency problems.
- Finally, we illustrate the feasibility and advantages of the expected FSPVM based on the proposed GERM framework and EVI by implementing a toy IPL for IMP [25] and a formal interpreter in Coq based on the GERM framework and EVI to simulate the situation that the four challenges have been overcome, and apply them to verify the properties of a simple program segment written in IMP.

We employ Coq in this work because it is one of the most highly regarded and widely employed proof assistants [12]. The work in this paper is the first step of our ongoing project to build a general semi-automatic formal verification FSPVM for verifying smart contracts operating on the blockchain platform easily and reliably which will overcome the four challenges mentioned above. Our intention is to submit this as an open source project after completion of the core work.

The remainder of this paper is structured as follows. Section 2 introduces the related work about studies on consistence, reusability and automation problems. Section 3 gives the basic notion and background of CHI. Section 4 introduces overall structure of the formal memory framework, and provides formal definitions of each of its components along with relevant proofs of their correct functionality. Section 5 elaborates on the basic concept and advantages of EVI. Section 6 emphasizes the feasibility of the excepted FSPVM based on the proposed

GERM framework and EVI by a simple instance. Finally, Section 7 presents preliminary conclusions and directions for future work.

2. Related work

Program verification using higher-order logic theorem proving is a very important theoretical field in computer science. Many researchers try to solve the consistence, reusability and automation problems from different aspects and develop new tools to contribute to this field. For consistence and reusability problems, one of the pretty standard and efficient methods is to formalize real world programming languages as a IPL and design a formal memory model as the state model. Since the late 1960's, a very large number of studies have focused on building memory models mathematically for program verification. Here, we present a brief discussion of the most significant studies that have inspired the present work. Norrish [33] and Hohmuth et al. [34] provided mechanized C/C++ semantics in HOL and PVS, respectively, which included low-level memory models. Tuch et al. [35] developed the first treatment of separation logic that unified the byte-level and logical views of memory in Isabelle/HOL. Appel and Blazy [32] later developed a mechanized separation logic for a C-based intermediate language in Coq. Manson et al. [29] developed a Java memory model. However, these works focus on specific domains and programming languages, and their formal memory models are deeply embedded in their framework, making them difficult to extend and modify for supporting different high-level specifications, which would enable the formalization of programs written by different high-level languages. Besides, most of them are individual researches for one or two problems on consistence, reusability or automation problems instead of considering them simultaneously. So they are hard to extend to solve these problems all.

In 2008, one of the milestones, CompCert project, appears which aims at compiler verification [16]. The team of CompCert formalizes an equivalent IPL called Clight for C language which is mechanized it in Coq. Besides, they develop a formal memory model for low-level imperative languages such as C and compiler intermediate languages. These works has served as the basis of some interesting and powerful program verification and analysis frameworks. Verified software toolchain (VST) [10] and deep specifications [11] are two representative projects which have been developed in conjunction with the IPL and formal memory model provided by CompCert. CertiKOS [7] is one of the most successful verification examples of them. In addition, seL4 [6] is another similar well-known project in recent years which is a fully verified microkernel that is considered to be the first OS kernel developed with an end-to-end proof of implementation correctness and security enforcement. However, these analysis frameworks suffer from the following main disadvantages.

- First, they still focus on specific domains and programming languages. Besides, although formal memory model of CompCert can be extended to support different high-level specifications, it is still deeply embedded in its framework, which is hard work for general users to analyze and modify to support their own researches based on it. Moreover, the functionality of their formal memory models depends on the details of their toolchains.
- Second, VST and deep specifications are very professional program verification tool chains and frameworks based on higher-order logic theorem proving for C programs. But they are unfriendly to general users. They have complex architectures, application programming interfaces (APIs) and tactics. These factors make the operations of these frameworks very difficult to be learned by general users. Besides, they are too heavyweight to be extended or modified by general users to solve special cases in their own researches..
- Third, the frameworks, such as deep specifications, need researchers to rebuild resource code of programs, and construct abstract layers manually. Even though, according to [11], verifiers should define specifications to prove the consistence between two relevant layers, the whole process is dependent on the experience of verifiers instead of a standard. Therefore, it still has the risk of consistence problem and it is impossible to become automatic using current automated theorem proving technologies.

Clearly, the first two disadvantages limit the ability of these analysis frameworks to handle special cases within their specific focus and severely restrict general users in the application of these frameworks, and the third disadvantage shows that current frameworks have not solved the consistence and automation problems completely. Finally, the formalization workload associated with these frameworks remains very heavy.

Compared with them, GERM provides almost the identical functionalities with CompCert, but its core design is much more lightweight, extensible and intuitive which can be easily implemented in Coq or similar proof assistants by a doctoral student who has the basic knowledge of the chosen proof assitant. Especially, GERM is not embedded in any other high-level frameworks. So it can generally server as the low-level formal memory model for arbitrary high-level program verification specifications. It also can serve as the basis of FSPVM which is our blueprint to solve the all three problems. Besides, we have implemented and verified GERM in Coq and extend it to support our ongoing project about formalizing Solidity [28] and the respective formal interpreter.

For automation problem, we note that symbolic execution is one of the best methods to improve the degree of automation. Unfortunately,

none of the above frameworks employs it in them. Some powerful automatic theorem proving assistants have been developed based on it, such as satisfiability modulo theories (SMT) or SMT-based theorem proving assistants [8]. But they do not readily support higher-order logic, such that the expressibility and provability of formalizations is limited. SMTCoq [30] is an interesting project that tries to combine SMT and Coq. However, it is not sufficiently mature to finish complex programs verification missions.

Aiming at these problems, EVI can combine higher-order theorem proving and symbolic execution directly, standardize the modeling and verifying process and make it possible to design full-automatically tactics to verify different formal models.

We hope that, in the future, our works might be useful in other contexts such as static analyzers and program provers and their formal verification.

## 3. The basic notion of CHI

In response to the work of H. B. Curry regarding programming language and proof theory, W. A. Howard privately circulated a manuscript in 1969, which was later formally published [22]. In this work, Howard pointed out that a correspondence exists between natural deduction and simply-typed lambda calculus, which established what is now denoted as CHI [23]. This work has served as the inspiration for many theorem proving assistants and functional programming languages (FPLs), such as Agda, Automath, Coq, Epigram, F#, F*, and Haskell, and is also a primary component forming the fundamental theory behind Cic. In brief, CHI proposes that a deep correspondence exists between the world of logic and the world of computation. This correspondence can be expressed according to three general principles. The first principle is given below.

*propositions as types* (Principle 1)

This principle describes an isomorphism between a given formal logic and a given programming language. At the surface, it says that, for each proposition in a formal logic, there is a corresponding type in an FPL, such as Coq (Gallina), and vice versa. The correspondence extends deeper, in that, for each proof of a given proposition, there is a program based on a lambda calculus of the corresponding type, and vice versa. This leads to the second principle given below.

*proofs as programs* (Principle 2)

Finally, the correspondence extends deeper still, in that each available means of simplifying a proof has a corresponding way of evaluating a program, and vice versa. This leads to the third principle given below.

*proofs as evaluation of programs* (Principle 3)

This theory is the basis of GERM and EVI. It is also an important reason for us to employ Coq to achieve these works.

## 4. Formal definition of the GERM framework

The GERM framework is designed and implemented based on the formal language denoted as Calculus of Inductive Construction (Cic) employed in Coq [15], which is well suited as a basis for high-level specifications in different formal models for program verification. For example, the use of Cic allows the GERM framework to be reused with different program verification formal models to store and generate intermediate states. In addition, this also serves as the basis of the EVI concept presented in Section 5.

The overall GERM framework structure is illustrated in Fig. 1. According to the figure, the GERM framework comprises two main components: a formal memory model in a trusted domain and assistant tools in a general domain. The formal memory model includes three levels from bottom to top: a formal memory space, low-level memory management operations, and basic memory management APIs. These levels are discussed in detail in Subsections 4.1.1–4.1.3, respectively. Assistant tools are employed in the GERM framework to obtain user requirements and generate dynamic specifications. Assistant tools are discussed in detail in Subsection 4.2.

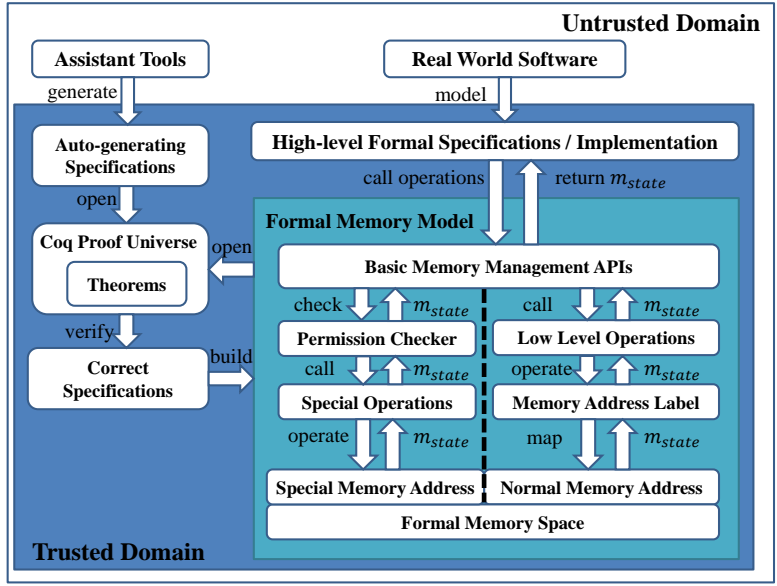

Figure 1. Architecture of the GERM framework

The workflow of the GERM framework can be defined in conjunction with Fig. 1 as follows. A user first sets initial requirements, such as memory size, and then the assistant tools generate the respective specifications. Next, the entire formal memory model is certified according to the correctness properties employed in Coq. In the Coq specification, the judgments of the dynamic semantics are encoded as mutually inductive predicates, and the functions are written in Gallina, which is a non-Turing complete language without halting problem. If the formal memory model satisfies all required properties, then the specific GERM framework has been constructed successfully. The user can then build a high-level formal model based on the generated GERM framework. The complete workload for constructing the GERM framework with 100 memory blocks is itemized in the Table 1.

Table 1. Workload statistics for constructing the GERM framework with 100 memory blocks.

|  | Objects | Lines in C++ | Lines in Coq |
|---|---|---|---|
| Formal Memory Space | 1 | 0 | 104 |
| Formal Memory Value | 1 | 0 | 11 |
| Formal Memory Operations | 122 | 0 | 2,289 |
| Assistant Tools | 6 | 423 | 0 |
| Correctness Lemmas | 23 | 0 | 531 |
| Total | 153 | 3,358 |  |

**4.1 Formal memory model**

**4.1.1 Formal memory space**

This level simulates a real world physical memory structure, and consists of formal memory blocks used to store information, and the formal memory addresses used to index the respective memory blocks. Because of the formal memory space definition employed in the GERM framework, we can define special memory addresses to index special memory blocks isolated from the normal memory block.

A number of interesting algorithms can be employed to abstract a formal memory space, such as tree structure mapping [16] or graphic mapping [17]. These algorithms have both advantages and disadvantages. For example, they are able to represent an infinite memory space. However, their specifications and formal structures are very complex and difficult to extend. Moreover, for an operation to modify a memory block, it must search all nodes one by one rather than modifying the block directly through its memory address. These disadvantages can complicate the verification process, and correspondingly increase the workload of the proof assistant.

To avoid these disadvantages, and to simulate the physical memory space more intuitively in the higher order logic system of Coq, we define the formal memory space architecture by enumerating the memory blocks using *Record* type. In details, the abstract syntax of *Record* type defined in Coq reference manual [14] is given in *Record* Syntax 1 and 2.

$record ::= Record\ ident\ [binders]\ [:sort] := [ident]\ \{\ [field_0\ ;\ \ldots;\ field_n]\ \}.$ (*Record* Syntax 1)

$$Record\ ident\ params : sort := ident_c \{ ident_0\ [binders_0]: term_0; \ldots ident_n\ [binders_n]: term_n \} \quad (Record\ \text{Syntax 2})$$

The square brackets "[ ]" means the term is optional. The identifier *ident* is the name of the defined record and *sort* is its type. The identifier $ident_c$ is the name of its constructor. The binders $binders_0$ to $binders_n$ are the quantifiers (such as $\forall$ and $\exists$) which is optional. The identifiers $ident_0$ to $ident_n$ are the names of fields and for all $binders_0$, $term_0$ to for all $binders_n$, $term_n$ their respective types. And it has some valuable features. First, according to the *Record* Syntax 2, each field must have an identifier. Second, in Coq, the identifiers are essentially abstract functions with type $[binders_i], ident \to term_i$, and the field identifiers are satisfied the Axiom 1 to 3.

**Axiom 1** (Identifier Uniqueness): For all field identifiers in the same record type, suppose for all $i, j \in \mathbb{N} \land i \neq j$, then $ident_i \neq ident_j$ holds.

**Axiom 2** (Bijection): As Relation 1, for all field identifiers are bound with a unique explicit field identifier and they satisfy the bijection function relationship.

$$identifiers \xleftrightarrow{bijection} fields \quad (\text{Relation 1})$$

**Axiom 3** (Field Access): For all field identifiers are bound with a unique explicit field identifier and a field can only be accessed by its binding identifier.

These three axioms have already been built in the trusted core of Coq by Coq development team. The type-checking mechanism will check the definitions in the higher-order logic system of Coq are whether satisfied the Axiom 1 to 3. And it will give the error message if some definitions do not pass the type-checking test. Here is a simple example in Table 2. When we define a *Record* type *example*, if the definition exist duplicate field names, then Coq will give the error message.

Table 2. An example of ill-formed definition of field identifiers checked by Coq.

| |
|---|
| Coq < Record example: Type := new {a : A; a : B}. |
| ----------------------------------------------------------- |
| Error: Objects have the same name |

After redefining the *Record* type *example* in Table. 3 to satisfy the Axiom 1, we construct a specific *Record* term *e* with type *example* to store logic values $v_a$ and $v_b$ which have type A and B. The only way, provided in Coq, to access the field values stored in *e* is invoking the respective binding field identifiers.

Table 3. An simple example of *Record* type about declaring, constructing and accessing.

| |
|---|
| Coq < Record example: Type := new {a : A; b : B}. |
| Coq < Definition e := new $v_a\ v_b$. |
| Coq < Eval simpl in (e.(a)). |
| ----------------------------------------------------------- |
| example is defined |
| a is defined |
| b is defined |
| e is defined |
| = $v_a$ : A |

Based on these useful features of *Record* type, the expected lightweight and intuitively formal memory space can be abstracted as following contents. In Relation 2 and 3, the field identifiers are specified to represent the memory address and the binding fields represent the corresponding memory blocks.

$$ident_i \xrightarrow{represents} address_i \quad (\text{Relation 2})$$
$$field_i \xrightarrow{represents} block_i \quad (\text{Relation 3})$$

An example of this method for the specification of 16 memory blocks is illustrated in Fig. 2. The left side of Fig. 2 represents the formal specification of memory space in Coq, and the right side is the real world physical memory space structure. In the formal specification, the memory address denoted by *address* is the field identifier of record type *memory*, and each field can record a term denoted as *value*. According to the definition, it is clear that each formal memory block can be abstracted as a Cartesian product $\langle m_{addr}, m_{value} \rangle : address * value$, where the metavariable $m_{addr}$ is an arbitrary memory address, and the metavariable $m_{value}$ is the *value* term stored in $m_{addr}$. Users not only can define normal memory addresses, but also can define addresses such as the address $m\_0xinit$ in Fig. 2 for special purposes, and, in this way, we can express isolation relationships between normal memory blocks and special memory blocks.

```
Record memory : Type := new {
    m_0xinit : value
    m_0x00000000 : value
    m_0x00000001 : value
    m_0x00000002 : value
    m_0x00000003 : value
    …
    m_0x0000000F: value
}
```

| Memory Address | Data Value |
|---|---|
| m_0xinit | value |
| m_0x00000000 | value |
| m_0x00000001 | value |
| m_0x00000002 | value |
| m_0x00000003 | value |
| … | … |
| m_0x0000000F | value |

Figure 2. Formal memory space, including the formal specification of memory space in Coq (left), and the real world physical memory space structure (right).

Formal definitions of the memory syntax and proofs of the uniqueness, singlenesss, and isolation relationships of memory addresses are provided as follows based on an example abstract syntax given as BNF-MEM-ADDR below.

Memory address:   $m_{addr}$ ::= special address | m_0x00000000 | … | m_0xFFFFFFFF  (BNF-MEM-ADDR)

**Definition 1** (*memory*; memory state, $m_{state}$): We use $m_{addr}{}^*$ to represent the set of memory addresses and $m_{value}{}^*$ to represent the set of *value* terms, so the formal memory specification can be defined as the rule MEM-SPACE below, and its constructor is denoted as *new*.

$$memory \equiv Record\langle m_{addr}{}^*, m_{value}{}^*\rangle \quad \text{(MEM-SPACE)}$$

In addition, we denote a logical unit of *memory* transferred between different layers over the entire trusted domain as the memory state. Each memory state records the current verification information, which is generated automatically by symbolic execution and reasoning. We use the metavariable $m_{state}$ to represent a memory state, and it is defined according to the rule MEM-STATE below.

$$m_{state}: memory \quad \text{(MEM-STATE)}$$

**Theorem 1** (Uniqueness): We define *a* and *a'* as two arbitrary formal memory address, and state that for all $i, j \in \mathbb{N} \wedge i \neq j, address_i \neq address_j$.

*Proof.* According to the Relation 2 and 3, *addresses* are field identifiers of *Record* type and memory blocks are the binding fields. So *addresses* and memory blocks of this formal memory space model are satisfied the Axiom 1 to 3. And we can get the result that

$$\forall\, i, j \in \mathbb{N} \;\wedge\; i \neq j, address_i \neq address_j,$$

Hence an arbitrary formal memory block has a unique memory address and can only be accessed by its binding *address*.

**Theorem 2** (Singleness): We define *t* as an arbitrary formal memory block, and state that *t* can only have a single memory address.

*Proof.* Similar to Theorem 1, we can get that *addresses* and memory blocks of this formal memory space model are satisfied the Axiom 1 to 3. Therefore, we can get the result that

$$addresses \xleftrightarrow{bijection} blocks.$$

And we have proven that there not exist two equivalent *addresses*. Hence an arbitrary formal memory block *t* has a single memory address.

**Corollary 1** (Isolation): Again defining *t* as an arbitrary formal memory block, we state that *t* can only be accessed by its respective address.

*Proof.* Based on Theorem 1 and 2, we can derive that *t* is both unique and singular. Hence it is obvious that *t* can only be accessed by its respective address. And Corollary 1 also can be proven by Axiom 3 directly.

According to the above discussion, *value* includes all memory information about a memory block, and *memory* is dependent on the specification of *value*. In the current memory model, meatavariable $m_{value}$ represents an individual *value* term that consists of input data, data reflecting the respective environment, and the state of the respective memory block. Here, the data environment and state of the respective memory block are defined only abstractly in Definition 2 because they are dependent on the specific verification environment and requirements.

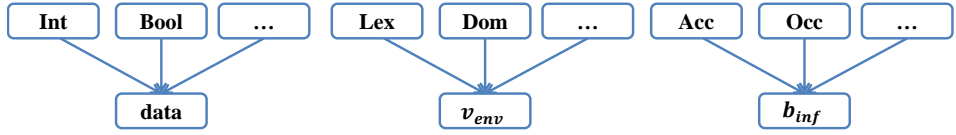

Figure 3. Expandability of the data framework, including datatypes (data), e.g., integers (Int) and Boolean (Bool) values, data environmental factors ($v_{env}$), e.g., the lexical scope (Lex) and lexical domains (Dom), and memory block state factors ($b_{inf}$), e.g., access authority (Acc) and occupation (Occ).

**Definition 2** (data environment; memory block state): The data environment variable *Env* is employed to formalize the context of respective *value* terms in the high-level programming language. In the current framework illustrated in Fig. 3, *Env* includes the lexical scope (Lex), lexical domains (Dom), *type* signatures, current context, inheritance, and super context. The variable $v_{env}$ represents an individual *Env* term, as follows.

$$v_{env}: Env \quad \text{(ENV-TERM)}$$

The memory block state variable *Blc* is employed to formalize the execution information of a memory block. In the current framework (Fig. 3), *Blc* includes the memory foot point, access authority (Acc), and occupation (Occ). The variable $b_{inf}$ represents an individual *Blc* term, as follows.

$$b_{inf}: Blc \quad \text{(BLC-TERM)}$$

Table 4. Basic datatypes employed in the formal memory space and their corresponding basic type inference rules of *value* constructors

$\boxed{value}$:

$$\frac{\mathcal{E} \vdash tt : unit \quad Undef : unit \to Env \to Blc \to value}{M,\mathcal{E} \vdash Undef\ tt\ v_{env}\ b_{inf} : value} \quad \text{(VALUE-UNDEF)}$$

$$\frac{\mathcal{E} \vdash oint : option\ int \quad Int : option\ int \to Env \to Blc \to value}{M,\mathcal{E} \vdash Int\ oint\ v_{env}\ b_{inf} : value} \quad \text{(VALUE INT)}$$

$$\frac{\mathcal{E} \vdash obool : option\ bool \quad Bool : option\ bool \to Env \to Blc \to value}{M,\mathcal{E} \vdash Bool\ obool\ v_{env}\ b_{inf} : value} \quad \text{(VALUE-BOOL)}$$

$$\frac{\mathcal{E} \vdash ofloat : option\ float \quad Float : option\ float \to Env \to Blc \to value}{M,\mathcal{E} \vdash Float\ ofloat\ v_{env}\ b_{inf} : value} \quad \text{(VALUE-FLOAT)}$$

$$\frac{\mathcal{E} \vdash ostring : option\ string \quad Strings : option\ string \to Env \to Blc \to value}{M,\mathcal{E} \vdash Strings\ ostring\ v_{env}\ b_{inf} : value} \quad \text{(VALUE-STRING)}$$

$$\frac{\Lambda \vdash initaddr : L_{address} \quad t : \tau \quad M,\mathcal{E},\Lambda \vdash v : value \quad \mathcal{E} \vdash blocksize : nat \quad Array : L_{address} \to \tau \to value \to nat \to Env \to Blc \to value}{M,\mathcal{E},\Lambda \vdash Array\ initaddr\ blocksize\ t\ v\ v_{env}\ b_{inf} : value} \quad \text{(VALUE-ARR)}$$

$$\frac{\Lambda \vdash oaddr : option\ L_{address} \quad vid : option\ L_{address} \to variable_{id} \quad Vid : variable_{id} \to Env \to Blc \to value}{M,\Lambda \vdash Vid\ (vid\ oaddr)\ v_{env}\ b_{inf} : value} \quad \text{(VALUE-PTR-VAR)}$$

$$\frac{\Lambda \vdash oaddr : option\ L_{address} \quad pid : option\ L_{address} \to parameter_{id} \quad Pid : parameter_{id} \to Env \to Blc \to value}{M,\Lambda \vdash Pid\ (pid\ oaddr)\ v_{env}\ b_{inf} : value} \quad \text{(VALUE-PTR-PAR)}$$

$$\frac{\Lambda \vdash oaddr : option\ L_{address} \quad fid : option\ L_{address} \to function_{id} \quad M,\mathcal{E},\Lambda \vdash opars : option\ (list\ value) \quad Fid : parameter_{id} \to option\ (list\ value) \to Env \to Blc \to value}{M,\mathcal{E},\Lambda \vdash Fid\ (fid\ oaddr) opars\ v_{env}\ b_{inf} : value} \quad \text{(VALUE-PTR-FUN)}$$

$$\frac{stt : statement \quad Stt : statement \to Env \to Blc \to value}{M \vdash Stt\ stt\ v_{env}\ b_{inf} :: value} \quad \text{(VALUE-STT)}$$

$$\frac{\Lambda \vdash name : L_{address} \quad members : struct_{mem} \quad Str_{type} : L_{address} \to struct_{mem} \to Env \to Blc \to value}{M,\Lambda \vdash Str_{type}\ name\ members\ v_{env}\ b_{inf} : value} \quad \text{(VALUE-STR-Type)}$$

$$\frac{\Lambda \vdash addr : address \quad M,\mathcal{E},\Lambda \vdash ovalues : option\ (list\ value) \quad Str : L_{address} \to option\ (list\ value) \to Env \to Blc \to value}{M,\mathcal{E},\Lambda \vdash Str\ addr\ ovalues\ v_{env}\ b_{inf} : value} \quad \text{(VALUE-STR)}$$

In actual practice, programs are translated into machine code by a compiler, and the execution information is stored in physical memory using binary code. However, the GERM framework focuses on the verification of high-level programs rather than their execution of on actual hardware. In addition, the use of formal machine code will reduce the readability of formal specifications and proving theorems, and increase the verification workload. Therefore, standard program execution information is stored by *value* directly in formal memory rather than translating that information into formal specifications of binary code. As such, *value* can be formalized as a tuple $\langle d, v_{env}, b_{inf} \rangle$, where, as shown in Fig. 3,

$d$ represents input data (data) of various types such as integer (Int) and Boolean (Bool), which are presented in detail at the end of this subsection. Due to the above definition, $d$ should be able to describe terms with different datatypes in the high-level programming language. In Cic, $d$, $v_{env}$ and $b_{inf}$ can be defined as inductive types, and different datatypes and memory properties are specified as different inductive type constructors of corresponding inductive types. In this way, the formal specifications of memory values are more readable, and can be transferred and extended easily by modifying the elements of the tuple, or by including additional inductive type constructors therein regarding data environment and/or memory block state freely. Furthermore, the use of this framework allows some common problems, such as memory overflow, to be found easily by type checking. Of course, if it becomes necessary to formalize the real world memory structure based on binary code, a user need only replace the current *value* definition with its corresponding binary definition.

According to the above discussion, we can formally define *value* in this framework according to Definition 3.

**Definition 3** (*value*; $m_{value}$): In the formal memory model of the present framework, the information recorded in an arbitrary *value* term, denoted as metavariable $m_{value}$, includes the input data, and respective data environment and memory block state information, where the sets of each are given by $d^*$, $v_{env}^*$, and $b_{inf}^*$, respectively. The formal specifications can be defined as the rule MEM-VALUE below.

$$value \equiv Inductive\ \langle d^*, v_{env}^*, b_{inf}^* \rangle \quad \text{(MEM-VALUE)}$$

$$m_{value}: value \quad \text{(MEM-VALUE-TERM)}$$

In current version of the proposed framework, *value* can record the 11 basic datatypes shown in Table. 4, including undefined (UNDEF), machine integer (INT), Boolean (BOOL), floating point (FLOAT) [18], string (STRING), array (ARR), pointers for variables (PTR-VAR), parameters (PTR-PAR), and functions (PTR-FUN) program statement (STT), and struct (with STR). As indicated by the table, most constructors are parameterized by the current *memory* information $M$, *label address* set $\Lambda$, and logic environment $\mathcal{E}$. We adopt the *unit* datatype to represent undefined, monad option data [19], which is the equivalent to the *maybe* datatype employed in Haskell [36]. The *unit* datatype describes the condition where a block is initialized but records no data value. In addition, the *struct* datatype has two inference rules, including VALUE-STR-Type, which is used to store the *struct* datatype declared by a programmer, and VALUE-STR, which is used to store the values of a variable declared by a *struct* datatype. In addition, the *struct* datatype inference rules include the *label address* variable $L_{address}$, which is a label of *address* that is presented in detail in the next subsection.

### 4.1.2 Low-level memory management operations

This level analyzes requests for high-level memory management operations, and interacts with the formal memory space to generate the resulting memory state for those operation requests. Finally, the operation requests are then executed at this level. Specifically, the interactions on this level involve two parts. The first part is illustrated at the right of Fig. 1, and represents the low-level operations for normal memory blocks. This interaction provides a set of address labels and low-level operations for normal memory blocks. The label address set satisfies a bijection relationship with the set of normal memory addresses, and is provided to facilitate the indexing and operation of normal memory blocks through low-level operations for basic APIs and high-level specifications. This represents a formal memory management layer designed in Coq to facilitate the interactions between the formal memory space and low-level management operations. In this way, the formal memory space is isolated from high-level specifications, and the special memory blocks are transparent to high-level specifications, which provides for safer and more effective operations management. The second part is illustrated at the left of Fig. 1, and represents the low-level operations for special memory blocks. If a high-level specification request passes the permission checking process, then the special memory block is controlled directly by the special operation.

In detail, this layer first provides a *label address* with a value that is given by the enumeration type variable $L_{address}$. An example abstract syntax of $L_{address}$ is defined as the rule BNF-LAB-ADDR below.

$$L_{address}: \quad a ::= \_0x00000000\ |\ \ldots\ |\ \_0xFFFFFFFF \quad \text{(BNF-LAB-ADDR)}$$

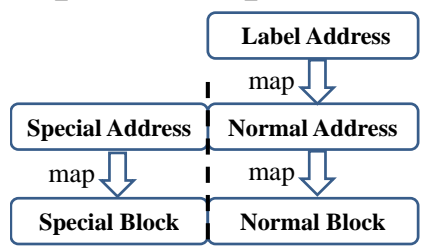

Figure 4. Mapping relationships of *label address*

Defining this kind of transitional type in Coq has two important benefits. First, as mentioned above, the field identifier (*address*) of a *Record* type actually represents a abstract function with type $memory \rightarrow value$ in Coq, which increases the degree of complexity of other high-level operation functions, and makes it more difficult to define and access a memory block. As such, the use of $L_{address}$ and can simplify the high-level execution specification. Nevertheless, we will continue to use *address* to represent $memory \rightarrow value$ in the remainder of this paper for simplicity of presentation. Second, based on the rule BNF-LAB-ADDR above, users can define some special memory block, such as *m_0xinit*, which would be transparent to high-level specifications because, as shown in Fig. 4, *label address* is a subset of *address*, and only maps to a normal memory block, which can be directly modified by high-level specifications. In this way, the transparency of the special memory block isolates the low-level memory block from the high-level specifications. The formal definition of $L_{address}$ is given in Definition 4.

**Definition 4** (*label address* type; $L_{address}$): In the low-level memory management layer, the *label address* type is a transitional type in Coq that is employed to provide a simple memory address identifier for operation functions, and to isolate the low-level formal memory space from high-level formal specifications. The value of the *label address* type is denoted as $L_{address}$, and the enumeration items of $L_{address}$ are denoted by $a^*$. The type definition is given in rules LAB-ADDR and LAB-ADDR-TERM below.

$$L_{address} \equiv Enum\ a^* \quad \text{(LAB-ADDR)}$$

Here, $a \in a^*: L_{address}$ and $m_{addr} \in m_{addr}^*: address$ can imply $\mathcal{M}ap\ a^* \subseteq m_{addr}^*$.

$$a: L_{address} \quad \text{(LAB-ADDR-TERM)}$$

Based on Definition 4, we can implement the basic formal memory operations using the FPL Gallina [14] in Coq. The present version of the formal memory model employs 20 verified memory management operations, including 14 basic APIs, which are summarized in Table 5, and three low-level operations and three special operations that are summarized in Table 6. These operations are discussed in detail in the following subsection based on the following definition.

**Definition 5** (initial data, $v_{init}$; initial memory space, $m_{init}$): The present formal framework employs a convention that defines the initial data of a memory block ($v_{init}$) as $(Undef\ tt\ v_{env}\ b_{inf})$, and its corresponding initial value of $m_{state}$ is defined as $m_{init}$.

Table 5. Basic memory management APIs employed in the formal memory model.

| Function | Description | Automatic |
|---|---|---|
| $\mathcal{M}ap_{L2m}$ | Map a to $m_{addr}$ | Yes |
| $\mathcal{M}ap_{m2L}$ | Map $m_{addr}$ to a | Yes |
| $\mathcal{M}ap_{L2nat}$ | Map a to $\mathbb{N}$ | Yes |
| $\mathcal{M}ap_{nat2L}$ | Map $\mathbb{N}$ to a | Yes |
| $read_{dir}$ | Read $m_{value}$ from a directly | No |
| $read_{chck}$ | Read $m_{value}$ from a after validation checking | No |
| $write_{dir}$ | Write $m_{value}$ at a directly | Yes |
| $write_{chck}$ | Write $m_{value}$ at a after validation checking | No |
| $address_{offset}$ | Offset address a to $a'$ | No |
| $address_{srch}$ | Search a specified memory block | No |
| $empty_{srch}$ | Search an empty memory block | No |
| allocate | Allocate a memory block | No |
| $free_{mem}$ | Free a specified memory block | No |
| $init_{mem}$ | Initialize the entire memory space | Yes |

Table 6. Low level operations employed in the formal memory model, where the first 3 are normal operations and the final 3 are special operations.

| Function | Description | Automatic |
|---|---|---|
| $read_{low}$ | Read $m_{value}$ stored at $m_{addr}$ directly | No |
| $write_{low}$ | Write $m_{value}$ at $m_{addr}$ directly | Yes |
| $infor_{chck}$ | Determine whether a memory block can be modified | No |
| $value_{dec}$ | Determine whether two $m_{value}$ entries are equal | No |
| $alloc_{chck}$ | Determine whether a memory block can be allocated | No |
| $set_{all}$ | Set the entire memory space according to $m_{value}$ | Yes |

**4.1.3 Formal memory management APIs**

This layer includes a set of simple, nonintrusive memory management APIs providing for implementing the high-level specifications designed by general users. The implementations of this layer are entirely independent from high-level specifications, and seamlessly implement the low-level memory management operations of the GERM framework, such that users need know nothing about these operations. Thus, any user who understands the basic grammar of Coq can modify or extend APIs easily according to specific requirements. Note that we use metavariable $\sigma$ to represent current memory state in formal definition of the operations. These operations can be broadly classified as *Map*, *Read*, *Write*, *Free*, *Allocate*, *Initialize*, and *Search*.

*Map*: The first four functions $\mathcal{M}ap_{L2m}$, $\mathcal{M}ap_{m2L}$, $\mathcal{M}ap_{L2nat}$, and $\mathcal{M}ap_{nat2L}$ represent the basic mapping operations in this formal memory model. These functions convert terms among nature numbers (*nat*), $L_{address}$, and *address*. Because $L_{address}$ is a subset of *nat* and *address*, these conversions represent a partial mapping between $L_{address}$ and *nat* or *address*. To accommodate a condition of mapping failure, an option was added into $\mathcal{M}ap_{m2L}$ and $\mathcal{M}ap_{nat2L}$, where, if the return value is *None*, the function failed to map a term from $L_{address}$ to *nat* or *address*, and the mapping was successful if the return value is *Some a*. The operational semantics of the *Map* operations are defined in Fig. 5. Here, we present only the lemma for the inversion property of the mapping functions. This is because the correctness of these functions is dependent on a specific definition of the mapping relation, which is abstract in this case, so that satisfying the inversion property means that every *label address* satisfying an arbitrary mapping relationship has a unique respective *address* and natural number, and vice versa. We can therefore assume under these conditions that the mapping operation in this model functions correctly.

$$\frac{\Lambda \vdash a: L_{address} \quad M \vdash m_{addr}: address \quad a \xrightarrow{map} m_{addr}}{\mathcal{M}ap_{L2m}: L_{address} \rightarrow address} \quad \text{(MAP-LM)} \qquad \frac{\Lambda \vdash a \ L_{address} \quad \mathcal{E} \vdash n: nat \quad a \xrightarrow{map} n}{\mathcal{E}, \Lambda \vdash \mathcal{M}ap_{L2nat}(a) \hookrightarrow n} \quad \text{(MAP-LN)}$$

$$\frac{\Lambda \vdash a :: L_{address} \quad M \vdash m_{addr}: address \quad a \xleftarrow{map} m_{addr}}{M, \Lambda \vdash \mathcal{M}ap_{m2L}(m_a) \hookrightarrow Some\ a} \quad \text{(MAP-ML-T)} \qquad \frac{\Lambda \vdash a :: L_{address} \quad M \vdash m_{addr}: address \quad a \nleftarrow m_{addr}}{M, \Lambda \vdash \mathcal{M}ap_{m2L}(m_a) \hookrightarrow None} \quad \text{(MAP-ML-F)}$$

$$\frac{\Lambda \vdash a :: L_{address} \quad \mathcal{E} \vdash n: nat \quad a \xleftarrow{map} n}{\mathcal{E}, \Lambda \vdash \mathcal{M}ap_{nat2L}(n) \hookrightarrow Some\ a} \quad \text{(MAP-NL-T)} \qquad \frac{\Lambda \vdash a: L_{address} \quad \mathcal{E} \vdash n: nat \quad a \nleftarrow n}{\mathcal{E}, \Lambda \vdash \mathcal{M}ap_{nat2L}(n) \hookrightarrow None} \quad \text{(MAP-NL-F)}$$

Figure 5. Operational semantics of *Map* operations

**Lemma 1** (inversion): Suppose for all $a: L_{address}$, the conditions $Some\ a = \mathcal{M}ap_{m2L}(\mathcal{M}ap_{L2m}(a))$ and $Some\ a = \mathcal{M}ap_{nat2L}(\mathcal{M}ap_{L2nat}(a))$ hold.

*Proof.* We split the conjunction as $Some\ a = \mathcal{M}ap_{m2L}(\mathcal{M}ap_{L2m}(a))$ and $Some\ a = \mathcal{M}ap_{nat2L}(\mathcal{M}ap_{L2nat}(a))$, and first prove the left part. 1) Because *label address* is a subset of *address* by Definition 4, all $a: L_{address}$ can map to a unique $m_a: address$ in any mapping relation ($\rightarrow$). Hence, through the rule MAP-LM (Fig. 5), $\mathcal{M}ap_{L2m}(a)$ can obtain a respective $m_a$. Obviously, $m_a$ can also map to $a$ though a reverse mapping relation ($\leftarrow$). Hence, through the rule MAP-ML-T (Fig. 5), $\mathcal{M}ap_{m2L}(m_a)$ can obtain $Some\ a$, which completes the proof for the left part. 2) The right part can be proven by the same process employed to prove the left part. Hence, the mapping operations satisfy the inversion property.

*Read*: The four read operation functions are $read_{low}$, $read_{dir}$, $Infor_{check}$, and $read_{chck}$. The low-level read operation is $read_{low}$, which is a special operation. This function is actually a redefinition of a Coq mechanism that has been introduced because each field identifier of *Record* type *memory* has the relation $memory \rightarrow value$, such that a term with *Record* type *memory* can access the value stored at a specified field by indexing the field identifier. For example, the construction *m.(m_0x00000003)* for a term *m* of type *memory* in Coq represents obtaining the value stored in the field indexed by the identifier *m_0x00000003* directly. Because the fundamental theory of Coq is a type of higher-order typed lambda calculus. So this is encapsulated by the lambda abstraction, abbreviated as term $\lambda$, as rule READ-LOW below.

$$read_{low} \equiv \Big(\lambda\, m::memory.\big(\lambda\, m_{addr}:address.(m.(m_{addr}))\big)\Big):value \quad \text{(READ-LOW)}$$

In addition, $read_{dir}$ employs the $read_{low}$ function and *Map* operations. The parameter of $read_{dir}$ is a term of $L_{address}$ type, and the term is translated into the parameter of $read_{low}$ by a *Map* operation. The function $Infor_{check}$ is a general low-level operation that is used for determining whether a memory block can be modified, and is also used to define other basic operations. Here, we only provide an abstract definition because the operation of $Infor_{check}$ is dependent on specific requirements and conditions. Specifically, for example, in the basic version of GERM, $Infor_{check}$ is employed to check access authority and type safety. However, when GERM is employed to support Ethereum verification, $Infor_{check}$ should add functions to check "*gas*" and "*balance*" [28] of smart contracts, such that its functionality is not generally verifiable. So we conclude it is correct. Finally, the function $read_{chck}$ is a combination of the above functions, and returns an option type value given by *Some a* if the operation is successful, and, otherwise, returns *None*. The operational semantics of the *Read* operations are defined in Fig. 6.

$$\frac{\Lambda \vdash a:L_{address} \quad \mathcal{E}\vdash M.(\mathcal{M}ap_{L2m}(a)) \hookrightarrow v \quad M,\mathcal{E},\Lambda \vdash \langle \sigma, read_{low}(M,\mathcal{M}ap_{L2m}(a))\rangle \hookrightarrow \langle \sigma',v\rangle \wedge \sigma \equiv \sigma'}{read_{dir}: memory \rightarrow L_{address} \rightarrow value \quad M,\mathcal{E},\Lambda \vdash \vdash \langle \sigma, read_{dir}(\sigma,a)\rangle \hookrightarrow \langle \sigma',v\rangle} \quad \text{(READ-DIR)}$$

$$\frac{\Lambda \vdash a::L_{address} \quad \mathcal{E}\vdash M.(\mathcal{M}ap_{L2m}(a)) \hookrightarrow v \quad M \vdash Infor_{check}(v_{env},b_{inf}) \hookrightarrow false}{read_{chck}: memory \rightarrow Env \rightarrow Blc \rightarrow L_{address} \rightarrow option\ value \quad M,\mathcal{E},\Lambda \vdash \langle \sigma, read_{chck}(\sigma,v_{env},b_{inf},a)\rangle \hookrightarrow \langle \sigma',false,None\rangle \wedge \sigma \equiv \sigma'} \quad \text{(READ-CHCK-FALSE)}$$

$$\frac{\Lambda \vdash a\ L_{address} \quad \mathcal{E}\vdash M.(\mathcal{M}ap_{L2m}(a)) \hookrightarrow v \quad M \vdash Infor_{check}(v_{env},b_{inf}) \hookrightarrow true}{read_{chck}: memory \rightarrow Env \rightarrow Blc \rightarrow L_{address} \rightarrow option\ value \quad M,\mathcal{E},\Lambda \vdash \langle \sigma, read_{chck}(\sigma,v_{env},b_{inf},a)\rangle \hookrightarrow \langle \sigma',true,Some(read_{dir}(\sigma',a))\rangle \wedge \sigma \equiv \sigma'} \quad \text{(READ-CHCK-TRUE)}$$

Figure 6. Operational semantics of *Read* operations

**Lemma 2** ($read_{low}$ correctness): Suppose for all $m_a:address$ and $m_{state}:memory$, the equality $read_{low}(m_{state},m_a) = m_{state}.(m_a)$ holds.

***Proof***. The rule READ-LOW is a redefinition of the Coq mechanism *M.(field_identifier)* using lambda abstraction. When $read_{low}$ applies $m_a$ and *m*, according to lambda application rules of Cic, we can derive rule READ-App.

$$read_{low}(m_{state},m_a) \equiv \Big(\lambda\, m:\ memory.\Big(\big(\lambda\, m_{addr}:address.(m.(m_{addr}))\big)m_a\Big)\Big)m_{state} \quad \text{(READ-App)}$$

Then we employ substitution rules of Cic to derive rule READ-Sub.

$$read_{low}(m_{state},m_a) \equiv \Big(\lambda\, m:\ memory.\Big(\big(\lambda\, m_{addr}:address.(m.(m_{addr}))\big)[m_{addr} \coloneqq m_a]\Big)\Big)[m \coloneqq m_{state}] \quad \text{(READ-Sub)}$$

Finally, we simplify rule Substitution can get $read_{low}(m_{state},m_a)$. Therefore, Lemma 2 is correct.

**Lemma 3** ($read_{dir}$ correctness): Suppose for all $a:L_{address}$, $m_a:address$, and $m:memory$ that, if $a$ maps to $m_a$, then $read_{dir}(m,a) = m.(m_a)$.

***Proof***. By applying the rule READ-DIR (Fig. 6), we can replace $read_{dir}(m,a)$ with $read_{low}(m,\mathcal{M}ap_{L2m}(a))$ (1). Lemmas 1 and 2 prove that $\mathcal{M}ap_{L2m}(a) \Rightarrow m_a$ (2) and $read_{low}(m,m_a) = m.(m_a)$ (3). Therefore, substituting (2) into (1) yields $read_{low}(m,m_a)$, which, according to (3), verifies that $read_{dir}(m,a) = m.(m_a)$ is true.

**Lemma 4** ($read_{chck}$ correctness): Suppose for all $a:L_{address}$, $m:memory$, $v_{env}::Env$, and $b_{inf}:Blc$ that, if $Infor_{check}(v_{env},b_{inf}) \Rightarrow false$, then $read_{chck}(m,v_{env},b_{inf},a) = None$; else, $read_{chck}(m,v_{env},b_{inf},a) = Some\,(m.(m_a))$.

***Proof***. When $Infor_{check}(v_{env},b_{inf})$ returns true, we can apply rule READ-CHCK-TRUE (Fig. 6) to replace $read_{chck}(m,v_{env},b_{inf},a)$ with $Some(read_{dir}(M',a))$ (1). Lemma 3 proves that $read_{dir}(m,a) = m.(m_a)$ (2). Therefore, substituting (2) into (1) yields $Some\,(m.(m_a))$.

Hence, $read_{chck}(m, v_{env}, b_{inf}, a) = Some(m.(m_a))$ is true. Otherwise, we can apply rule READ-CHCK-FALSE (Fig. 6) to replace $read_{chck}(m, v_{env}, b_{inf}, a)$ with *None*. Hence, $read_{chck}(m, v_{env}, b_{inf}, a) = None$ is true.

*Write*: These operations include the special operation $write_{low}$, and the two basic APIs $write_{dir}$ and $write_{chck}$. Actually, $write_{low}$ represent a set of low-level operations that are defined using a Coq mechanism. In Coq, a new term of a *Record* type can only be constructed using its respective constructor. For example, to write a value $v$ into a memory block indexed by $m\_0x00000003$, we must use the constructor *new* defined in Definition 1 to generate a new memory state $m_{state} = (new\ (m.(m_0)\ ...\ m.(m_2)\ v\ ...)$. According to lambda calculus abstraction, we present this mechanism as a set of low-level operations $write_{low_i}$ ($i \in \mathbb{N}$), which are defined by the rule WRITE-LOW below. In addition, the rule MAP-RE below represents the mapping of a memory address to its corresponding $write_{low_i}$ which is satisfied bijection relationship, and $write_{low}$ taking $m_{addr_i}$ as parameter to employ the respective $write_{low_i}$.

$$write_{low_i} \equiv \left(\lambda\ v : value. \left(\lambda\ m : memory. \left(new\ (m.(m_{addr_0})\ ...\ m.(m_{addr_{i-1}})\ [v/m.(m_{addr_i})]\ ...)\right)\right)\right) : memory$$

(WRITE-LOW)

$$m_{addr_i} \xleftrightarrow{bijection} write_{low_i}$$ (MAP-RE)

The function $write_{dir}$ employs $write_{low}$. The parameter of $write_{dir}$ is a term of type $L_{address}$ that is translated into the parameter of $write_{low}$ by a *Map* operation. Finally, the function $write_{chck}$ is a combination of the above functions and $Infor_{check}$. If $Infor_{check}$ returns a false option, $m_{state}$ will not be changed; otherwise, it will generate a new $m'_{state}$. The operational semantics of the *Write* operations are defined in Fig. 7.

$$\frac{M,\mathcal{E},\Lambda \vdash a, a' :: L_{address} \quad M,\mathcal{E},\Lambda \vdash v :: value \quad a' \in \bar{a} \wedge a' \neq a}{M,\mathcal{E},\Lambda \vdash \langle\sigma, write_{low}(\sigma, Map_{L2m}(a), v)\rangle \hookrightarrow \langle\sigma', \sigma'\rangle} \\ \frac{M,\mathcal{E},\Lambda \vdash read_{dir}(\sigma', a) \equiv v \wedge read_{dir}(\sigma', a') \equiv read_{dir}(\sigma, a')}{write_{dir} : memory \to L_{address} \to value \to memory} \\ \overline{M,\mathcal{E},\Lambda \vdash \langle\sigma, write_{dir}(\sigma, a, v)\rangle \hookrightarrow \langle\sigma', \sigma'\rangle}$$ (WRITE-DIR)

$$\frac{\Lambda \vdash a :: L_{address} \quad M,\mathcal{E},\Lambda \vdash v : value}{M,\mathcal{E},\Lambda \vdash Infor_{check}(v_{env}, b_{inf}) \hookrightarrow false} \\ \frac{write_{chck} : memory \to Env \to b_{inf} \to L_{address} \to value \to memory}{M,\mathcal{E},\Lambda \vdash \langle\sigma, write_{chck}(\sigma, v_{env}, b_{inf}, a, v)\rangle \hookrightarrow \langle\sigma', false, \sigma'\rangle \wedge \sigma \equiv \sigma'}$$ (WRITE-CHCK-FALSE)

$$\frac{\Lambda \vdash a :: L_{address} \quad M,\mathcal{E},\Lambda \vdash v :: value}{M,\mathcal{E},\Lambda \vdash Infor_{check}(v_{env}, b_{inf}) \hookrightarrow true} \\ \frac{write_{chck} : memory \to Env \to b_{inf} \to L_{address} \to value \to memory}{M,\mathcal{E},\Lambda \vdash \langle\sigma, write_{chck}(\sigma, v_{env}, b_{inf}, a, v)\rangle \hookrightarrow \langle\sigma', true, write_{dir}(\sigma', a, v)\rangle \wedge \sigma \equiv \sigma'}$$ (WRITE-CHCK-TRUE)

Figure 7. Operational semantics of *Write* operations

**Lemma 5** ($write_{low}$ correctness): Suppose for all $i, j \in \mathbb{N} \wedge i \neq j$, $m_{addr_i}, m_{addr_j}$ : *address*, $m_{state}$ : *memory*, and $v_{new}$ : *value* that the conjunction $\left(write_{low}(m_{state}, m_{addr_i}, v_{new})\right).(m_{addr_i}) = v_{new} \wedge \left(write_{low}(m_{state}, m_{addr_i}, v_{new})\right).(m_{addr_j}) = m_{state}.(m_{addr_j})$ holds.

*Proof.* First, we destruct this conjunction into two sub-goals that $\left(write_{low}(m_{state}, m_{addr_i}, v_{new})\right).(m_{addr_i})$ (1) and $\left(write_{low}(m_{state}, m_{addr_i}, v_{new})\right).(m_{addr_j}) = m_{state}.(m_{addr_j})$ (2). For (1), According to the rule MAP-RE, $write_{low}(m_{state}, m_a, v_{new})$ can be replaced by $write_{low_i}(m_{state}, v_{new})$. Then, according to Cic application rules and the rule WRITE-LOW above, $write_{low_i}(m_{state}, v_{new})$ is replaced by rule WRITE-App.

$$write_{low_i}(m_{state}, v_{new}) \equiv \left(\lambda\ v : value. \left(\left(\lambda\ m : memory. \left(new\ (m.(m_{addr_0})\ ...\ [v/m.(m_{addr_i})]\ ...)\right)\right) m_{state}\right)\right) v_{new}$$

(WRITE-App)

Then we employ substitution rules of Cic to derive rule WRITE-Sub.

$$write_{low_i}(m_{state}, v_{new}) \equiv \left(\lambda\ v : value. \left(\left(\lambda\ m : memory. \left(new\ (m.(m_{addr_0})\ ...\ [v/m.(m_{addr_i})]\ ...)\right)\right)[m := m_{state}]\right)\right)[v := v_{new}]$$

(WRITE-Sub)

Finally, we can simplify WRITE-Sub as WRITE-Sub' which is a new memory state.

$$write_{low_i}(m_{state}, v_{new}) \equiv new\ (m_{state}.(m_{addr_0})\ ...\ [v_{new}/m.(m_{addr_i})]\ ...)$$ (WRITE-Sub')

Therefore, $(write_{low}(m_{state}, m_{addr_i}, v_{new})).(m_{addr_i}) \equiv new\ (m_{state}.(m_{addr_0})\ ...\ [v_{new}/m.(m_{addr_i})]\ ...\ ).(m_{addr_i}) = v_{new}$. Hence, (1) is correct. For (2), according to WRITE-Sub', it is obvious that for all $j \in \mathbb{N} \wedge i \neq j$, $(write_{low}(m_{state}, m_{addr_i}, v_{new})).(m_{addr_j}) \equiv new\ (m_{state}.(m_{addr_0})\ ...\ [v_{new}/m.(m_{addr_i})]\ ...\ ).(m_{addr_j}) = m_{state}.(m_{addr_j})$. Therefore, (2) is also correct. Thus, Lemma 5 is proven.

**Lemma 6** ($write_{dir}$ correctness): Suppose for all $i, j \in \mathbb{N} \wedge i \neq j$, $a_i, a_j : L_{address}$, $m_{state} : memory$, and $v_{new} : value$ that the conjunction $(write_{dir}(m_{state}, a_i, v_{new})).(m_{addr_i}) = v_{new} \wedge (write_{dir}(m_{state}, a_j, v_{new})).(m_{addr_j}) = m_{state}.(m_{addr_j})$ holds.

*Proof.* By applying the rule WRITE-DIR in Fig. 7 and MAP-LM in Fig. 5, we can replace this conjunction with $(write_{low}(m_{state}, m_{addr_i}, v_{new})).(m_{addr_i}) = v_{new} \wedge (write_{low}(m_{state}, m_{addr_i}, v_{new})).(m_{addr_j}) = m_{state}.(m_{addr_j})$. Then, we can apply Lemma 5 to prove it directly.

**Lemma 7** ($write_{chck}$ correctness): Suppose for all $a : L_{address}$, $m : memory$, $v_{env} :: Env$, and $b_{inf} : Blc$ that, if $a$ maps to $m_a$ and $Infor_{check}(v_{env}, b_{inf}) \hookrightarrow false$, then $write_{chck}(m, v_{env}, b_{inf}, a, v) = m$; else, $write_{chck}(m, v_{env}, b_{inf}, a, v) = write_{dir}(m, a, v)$.

*Proof.* By applying the rules WRITE-CHCK-FALSE and WRITE-CHCK-TRUE (Fig. 7), we can replace $write_{chck}$ with $m$ or $write_{dir}$, and prove the equalities directly.

$$\frac{\begin{array}{c} \Lambda \vdash a\ L_{address} \quad M \vdash m_a : address \quad \mathcal{E} \vdash offset : nat \\ f_{off} : nat \to nat \to nat \\ address_{offset} : L_{address} \to (nat \to nat \to nat) \to nat \to option\ L_{address} \end{array}}{M, \mathcal{E}, \Lambda \vdash address_{offset}(a, f_{off}, offset) \hookrightarrow Map_{nat2L}(f_{off}(Map_{L2nat}(a), offset))}\ \text{(ADDR-OFF)}$$

$$\frac{\begin{array}{c} A : Type \quad \Lambda \vdash a_{current} : L_{address} \\ filter : A \to bool \quad condition : A \\ filter(condition)\ true \\ address_{srch} : memory \to L_{address} \to (A \to bool) \to option\ L_{address} \end{array}}{M, \mathcal{E}, \Lambda \vdash \langle \sigma, address_{srch}(\sigma, a_{current}, filter) \rangle \hookrightarrow \langle \sigma', Some(a_{current}) \rangle \wedge \sigma \equiv \sigma'}\ \text{(ADDR-SER-T)}$$

$$\frac{\begin{array}{c} A : Type \quad \Lambda \vdash a_{current} : L_{address} \\ filter : A \to bool \quad condition : A \\ filter(condition) \hookrightarrow true \\ M, \mathcal{E}, \Lambda \vdash Map_{L2nat}(address_{offset}(Map_{L2nat}(a_{current}), plus, 1)) \hookrightarrow Some\ a_{next} \\ address_{srch} : memory \to L_{address} \to (A \to bool) \to option\ L_{address} \end{array}}{M, \mathcal{E}, \Lambda \vdash \langle \sigma, address_{srch}(\sigma, a_{current}, filter) \rangle \hookrightarrow \langle \sigma', address_{srch}(\sigma', a_{next}, filter) \rangle \wedge \sigma \equiv \sigma'}\ \text{(ADDR-SER-NE)}$$

$$\frac{\begin{array}{c} A : Type \quad \Lambda \vdash a_{current} : L_{address} \\ filter : A \to bool \quad condition : A \\ filter(condition) \hookrightarrow false \\ M, \mathcal{E}, \Lambda \vdash Map_{L2nat}(address_{offset}(Map_{L2nat}(a_{current}), plus, 1)) \hookrightarrow None \\ address_{srch} : memory \to L_{address} \to (A \to bool) \to option\ L_{address} \end{array}}{M, \mathcal{E}, \Lambda \vdash \langle \sigma, empty_{srch}(\sigma, a_{current}) \rangle \hookrightarrow \langle \sigma', None \rangle \wedge \sigma \equiv \sigma'}\ \text{(ADDR-SER-NO)}$$

Figure 8. Operational semantics of *Search* operations

*Search*: These are also essential basic operations that include functions $address_{offset}$, $address_{srch}$, $value_{dec}$, and $empty_{srch}$. The function $address_{offset}$ is applied for shifting a current *label address* to another *label address* according to a specified offset. It is defined as a higher-order function that takes a basic *label address*, a offset function and an arbitrary offset as parameters to accommodate different offset conditions. We should note that $address_{offset}$ is not always successful. Because according to *memory* definition, memory space has a fixed size, the result returned by $address_{offset}$ may over the range of *address* which is invalid. In order to deal this problem, we employ $Map_{nat2L}$ to check the return value of offset function, and $address_{offset}$ returns an option type value given by *Some label address*, if $Map_{nat2L}$ is successful, and, otherwise, returns *None*. The function $address_{srch}$ is employed in conjunction with $address_{offset}$ to search a specified memory block that satisfies a filter condition. The function $value_{dec}$ returns a binary sum datatype [20] defined by the rule VAL-DEC below. This rule can be proven easily in Coq using the *decide equality* tactic [14].

$$value_{dec} : (\forall v_0\ v_1 :: value, \{v_0 = v_1\} + \{v_0 \neq v_1\})\ \text{(VAL-DEC)}$$

Finally, $empty_{srch}$ is a special case of $address_{srch}$, where the filter is specified as $value_{dec}(v_{init}) :: (\forall v_1, \{v_{init} = v_1\} + \{v_{init} \neq v_1\})$. The correctness of $empty_{srch}$ is confirmed according to the correctness of $value_{dec}$ discussed above and the correctness of $address_{srch}$ presented by Lemma 9 below. The operational semantics of the *Search* operations are defined in Fig. 8.

**Lemma 8** ($address_{offset}$ correctness): Suppose for all $m_a$ : address, $m$ : memory, $v$ : value, $n$ : nat, and $f_{off}$ : nat → nat → nat that the equality $\mathcal{M}ap_{nat2L}(f_{off}(\mathcal{M}ap_{L2nat}(a), n)) = address_{offset}(a, f_{off}, n)$ holds.

*Proof.* The correctness of $\mathcal{M}ap_{nat2L}$ and $\mathcal{M}ap_{L2nat}$ have been proven by Lemma 1, so that Lemma 8 can be proven directly by applying the rule ADDR-OFF (Fig. 8).

**Lemma 9** ($address_{srch}$ correctness): Suppose for all $\{A : Type\}$, $filter : A \rightarrow bool$, $condition : A$, $a : L_{address}$, and $m : memory$ that, if $filter(condition) \hookrightarrow true$, then $address_{srch}(m, a, filter) = Some(a_{current})$ (1); else, if $filter(condition) \hookrightarrow false$ and the next address produced by $address_{offset}$ fails, then $address_{srch}(m, a, filter) = None$ (2); else, search the next indexed memory block recursively (3).

*Proof.* Based on Lemmas 1 and 8, cases (1), (2), and (3) above can be proven by applying rules (Fig. 8) ADDR-SER-T, ADDR-SER-NE, and ADDR-SER-NO, respectively.

*Allocate*: These operations are basic and essential APIs denoted by the functions $allocate$ and $alloc_{chck}$. The low-level operation $alloc_{chck}$ is used to determine whether the information of the current memory block, including various factors such as authority and occupation, satisfies the condition for allocation. Here, we only provide an abstract definition of $alloc_{chck}$ because its operation is dependent on specific requirements and conditions, such that its functionality is not generally verifiable. The function $allocate$ is a special case of $address_{srch}$, where the filter of $address_{srch}$ is specified using $alloc_{chck}$. Assuming the correctness of $alloc_{chck}$, in conjunction with the correctness of $address_{srch}$ proven in Lemma 9, we can conclude that the functionality of $allocate$ is also correct. The operational semantics of the *Allocate* operation are defined in Fig. 9.

$$\frac{\Lambda \vdash a\ L_{address} \quad M,\mathcal{E},\Lambda \vdash alloc_{chck}(read_{dir}(\sigma,a)) \hookrightarrow false}{M,\mathcal{E},\Lambda \vdash \langle \sigma, allocate(a,\sigma)\rangle \hookrightarrow \langle \sigma', address_{srch}(\sigma,a,alloc_{chck})\rangle \wedge \sigma \equiv \sigma'}\text{ (ALLOC)}$$

$$allocate : address \rightarrow memory \rightarrow option$$

Figure 9. Operational semantics of *Allocate* operations

*Free*: This operation is represented by the function $free_{mem}$, which is a special case of $write_{dir}$, where the input value $v$ is specified as $v_{init}$. Because the correctness of $write_{dir}$ has been proven in Lemma 6., the functionality of $free_{mem}$ is obviously correct. The operational semantics of the *Free* operation are defined in Fig. 10.

$$\frac{\Lambda \vdash a : L_{address}}{M,\mathcal{E},\Lambda \vdash \langle \sigma, free_{mem}(a,\sigma)\rangle \hookrightarrow \langle \sigma', write_{dir}(\sigma',a,v_{init})\rangle \wedge \sigma \equiv \sigma'}\text{ (FREE)}$$

$$free_{mem} : L_{address} \rightarrow memory \rightarrow memory$$

Figure 10. Operational semantics of *Free* operations

*Initialize*: These operations are employed to initialize the formal memory space prior to executing the verification process. These operations include the functions $set_{all}$ and $init_{mem}$, where $set_{all}$ is a special operation that implements a special case of $write_{low_i}$. The function of $set_{all}$ is to call all $write_{low_i}$ and modify the entire memory block using the value $v$. The correctness of every $write_{low_i}$ has been proven in Lemma 5, so the functionality of $set_{all}$ is obviously correct. The function $init_{mem}$ is a special case of $set_{all}$, where $v$ is specified as $v_{init}$. Thus, its functionality is also assured of being correct. The operational semantics of the *Initialize* operations are defined in Fig. 11.

$$\frac{M,\mathcal{E},\Lambda \vdash set_{all}(\sigma,v_{init}) \hookrightarrow \sigma_{init}}{M,\mathcal{E},\Lambda \vdash \langle \sigma, init_{mem}(\sigma)\rangle \Rightarrow \langle \sigma', \sigma_{init}\rangle \wedge \sigma_{init} \equiv \sigma'}\text{ (INIT)}$$

$$init_{mem} : memory \rightarrow memory$$

Figure 11. Operational semantics of *Initialize* operations

**4.2 Assistant tools**

According to Definition 1, the formal memory space and its respective addresses are defined by the *Record* type. This type is represented rigorously in the formal Cic language, and is a special inductive datatype with only a single constructor of type *Sort* [14]. This ensures that it is impossible to modify a *Record* type dynamically in Coq after it has been defined, which is analogous to the impossibility of dynamically changing the size of physical fixed-size memory hardware. Therefore, operations that depend on a specific memory address, such as *Map* and *Write* operations, also cannot modify memory space after it has been defined. This places considerable importance on the definition of formal memory spaces. However, defining a specific formal memory space by manually enumerating memory blocks and corresponding operations one by one can be an exceedingly tedious activity. Fortunately, Because of the enumeration process, the definitions of discussed formal memory space and related operations have fixed rules. For example, when we enumerate a new *address* in *memory* such as *m_0x00000003*, we should add *_0x00000003* into $L_{address}$. Then we need to update the mapping relationships and related operations mentioned above. The enumeration process of adding new *addresses* in *memory* is identical and the relevant operations are similar to each other. Therefore, these can be easily generated recursively by a simple program written in a high-level programming language such as Java or C++. This automatic definition of a memory space according to specific requirements is the basis of assistant tools. The formal definition of assistant tools is given below as the rule ASSIST-TOOL.

$$\Gamma, \mathcal{R} \vdash Tools\ r \hookrightarrow specifications \xrightarrow{yields} .v\ files \quad (\text{ASSIST-TOOL}).$$

Here, assistant tools function within a verification context $\Gamma$ with requirements $\mathcal{R}$, and employ a specific user requirement $r :: \mathcal{R}$ as parameters, which include special memory blocks and the size of normal memory blocks. Assistant tools then generate the respective formal specifications and export them as *.v files* that can be loaded in Coq directly. These formal specifications are denoted as dynamic specifications, and are based on Definitions 1 and 4. In addition, assistant tools employ those operations listed in Tables 2 and 3 that are given as automatic. The results in *.v files* generated by assistant tools are written using Gallina syntax, which can be executed and verified in Coq directly. According to the proofs given in the previous subsection, the results of assistant tools can be assumed to include no ill-formed definitions and to satisfy all specifications as long as the results pass the Coq type-checking mechanism. Finally, although these assistant tools are implemented in the general domain using general-purpose programming languages, the relation between the assistant tools and the respective results satisfies the non-aftereffect property, as illustrated in Fig. 12 [21]. As such, the verified results are not influenced by the assistant tools implementation.

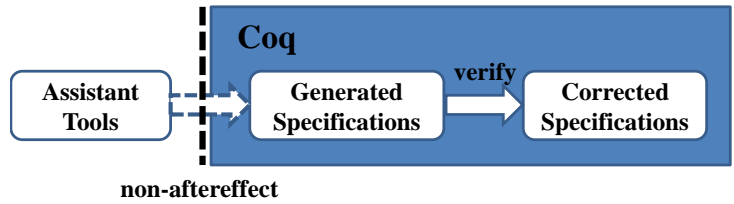

Figure 12. Workflow of assistant tools illustrating the non-aftereffect property

**4.3 Comparing with the formal memory model of CompCert**

Hereto, GERM framework is constructed completely. In order to further explain the value of GERM, we compare GERM with memory model of CompCert (MMoCC). Because CompCert is one of most successful project about program verification and it has been used in many other projects about higher-order theorem proving. Furthermore, MMoCC is also implemented in Coq.

Table 7. The comparison about basic features between the memory model of CompCert and GERM.

| Features | MMoCC | GERM |
| --- | --- | --- |
| Size | Dynamic and infinite | Static and finite |
| Weight | Heavyweight | Lightweight |
| Basic operations | alloc, free, load, store | alloc, free, read, write, initialize |
| Verification | Verified | Verified |
| Range of application | Low-level imperative languages | Generic |
| Pointer arithmetic | Support | Support |
| Embedment | Embedded | Non-embedded |

We compare them from the features of size, weight, basic operations, verification, range of application, pointer arithmetic and embedment. As illustrated in Table. 7, most of the features of MMoCC and GERM are identical. The differences between them are size, weight, range of application and embedment. The size of MMoCC is infinite, compared with finite size of GERM. But as mentioned in subsection 4, we have implemented the assistant tools to remedy this limitation, and the assistant tools of GERM can generate sufficient size automatically. Because MMoCC is specialized to support formal specifications of low-level imperative languages and compiler intermediate languages rather than aiming to supporting arbitrary specifications as GERM, MMoCC is embedded in CompCert framework rather than being independent as GERM. Finally, the weight of GERM is more lightweight than MMoCC. These features not only make GERM have almost identical functionalities with MMoCC, but make it friendlier to general users than MMoCC. As mentioned above sections, general users can redefine or extend GERM to support their own researches using basic knowledge of Coq. So GERM has huge potential and range of application in program verification field.

## 5. EVI

The concept of EVI is proposed herein to increase the degree to which the process of program verification is conducted automatically by combining higher-order logic theorem proving and symbolic execution. EVI includes three key components: a general formal memory model, a general-purpose IPL, and a respective formal interpreter. A general formal memory model such as GERM supports a basic formal system for constructing a logic-based operating environment corresponding to the real world operating environment of hardware, and serves as the basis for the general-purpose IPL and respective formal interpreter, which are used to model, execute, and verify programs automatically.

**5.1 Conceptual basis of EVI**

The deep correspondences make CHI very useful for unifying formal proofs and program computation. However, most mainstream general-purpose programming languages (GPLs) employed in the real world are not designed based on lambda calculus and cannot be analyzed in the higher-order logic environment. The programs written using these languages are very difficult or even impossible to verify directly and automatically using CHI. This forms the basis for the present development of EVI. To avoid ambiguity in the following discussion of EVI, we use *program* to represent programs that are written in an FPL based on CHI, and *RWprogram* to represent real world programs that are written in a GPL. In addition, we redefine metavariable $\mathcal{E}$ to represent higher-order logic environment which supports CHI.

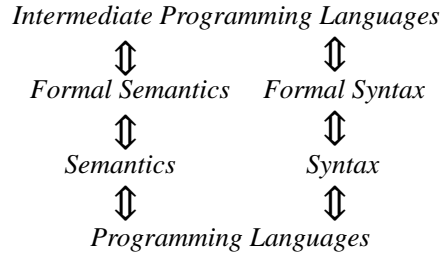

Figure 13. Formalization of a general-purpose programming language to obtain a intermediate programming language

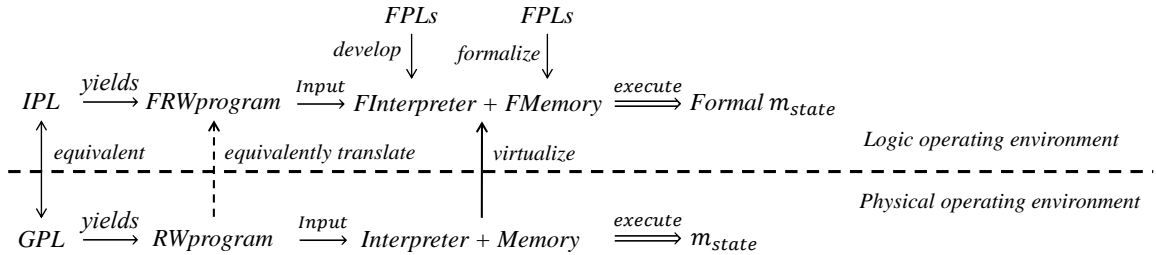

Figure 14. Equivalence between real world program (*RWprogram*) execution and execution in a logic environment

For the present development of EVI, we first note that, due to the equivalence between lambda calculus and Turing machine formalisms [24], logic theories based on lambda calculus are sufficiently powerful to formalize and mechanize the syntax and semantics of any GPL from beginning to end into $\mathcal{E}$ with the help of FPLs provided by $\mathcal{E}$, and thereby to obtain the corresponding IPL which can be analyzed in $\mathcal{E}$ directly, as illustrated in Fig. 13. Actually, there exist many related works, such as [40], to discuss how to build an IPL for a specific GPL. Because of the equivalent syntax and semantics, IPL is equivalent with GPL, so it can be implied that the formal *RWprogram* (*FRWprogram*) rewritten by IPL is also equivalent with *RWprogram* as Relation 4. In this way, when a *FRWprogram* is analyzed in $\mathcal{E}$ is equivalent that *RWprogram* is analyzed in $\mathcal{E}$.

$$GPL \equiv IPL \supset RWprogram \equiv FRWprogram \quad \text{(Relation 4)}$$

However, if we want to verify a *FRWprogram* using CHI, it should be executable in $\mathcal{E}$. In real world, an *RWprogram* is executed with the help of corresponding interpreters or compilers in a physical operating environment. Compared with physical operating environment, the combination of a formal general memory model such as GERM and its $\mathcal{E}$ have already virtualized a minimal higher-order logic operating environment. Therefore, although an *FRWprogram* cannot execute directly in $\mathcal{E}$, such as Coq, we can implement a formal interpreter (*FInterpreter*) using FPLs based on the higher-order logic operating environment that follows the formal syntax and semantics of the corresponding IPL to simulate the execution process of *RWprogram* in real world and interpret the *RWprogram* so that it can be symbolically executed in $\mathcal{E}$ direcly with the same process as is conducted in the real world. This process is illustrated in Fig. 14. Besides, the formal interpreter is developed using FPLs. Thus, the formal interpreter is a *program* with the following abstract definition:

$$formal\ interpreter: memory \to FRWprogram \to memory.$$

It means the formal interpreter takes a formal memory state defined by the general formal memory model and a *FRWprogram* as parameters and follows the formal semantics of IPL to execute (same as evaluating) *FRWprogram* and yields a new formal memory state. Accordingly, we can conclude that the execution of a formal *FRWprogram* in the formal interpreter corresponds to an evaluation of a *program*. And according to the Relation 4, we can also imply that it is equivalent with the symbolic execution of a formal *RWprogram*. Therefore, through Principles 2 and 3, we can derive Corollary 1 as follows.

*proofs as evaluation of programs as execution of programs* (Corollary 1)

While the above discussion demonstrates the isomorphism of *RWprogram* execution and proofs, we note that the process of proving *propositions* in present $\mathcal{E}$s is to construct equivalent propositions as its *proofs* [23]. According to Principle 1 to 3, it is isomorphic to constructing terms with excepted types by evaluating *programs*. Because of Corollary 1, we can also imply that it is also isomorphic to constructing terms with excepted types by executing *programs*. Obviously, the proofs cannot guarantee the correctness of properties excepted in *RWprogram*, such as functional correctness or security properties. For example, as illustrated in Fig. 15, the excepted correct memory state is $m_1$, but the final memory state evaluated by formal interpreter is $m_2$. Here we assume *FInterpreter* is correct (Because *FInterpreter* is a *program* developed using FPLs and can be directly verified its properties in $\mathcal{E}$.), so we can intuitively conclude that *FRWprogram* is not satisfied expected properties or functions, and it exists errors. But $m_1$ and $m_2$ both can be used as the *proofs* of *memory* type and the *proofs* cannot find out that *FRWprogram* exists errors.

$$FInterpreter(m_{init}, FRWprogram) \xrightarrow{exceptes} m_1 : memory \xrightarrow{prove} memory$$

$$FInterpreter(m_{init}, FRWprogram) \xrightarrow{evaluates} m_2 : memory \xrightarrow{prove} memory$$

Figure 15. Example of useless proofs

$$\begin{array}{ccc} proofs & \backsim & evaluation\ of\ programs \\ properties \Updownarrow & & \Updownarrow interpreter \\ verifications & \backsim & execution\ of\ RWprograms \end{array}$$

Figure 16. Isomorphism between verification and *RWprogram* execution

To solve this problem, we should strengthen the *propositions* according to the functional correctness requirements and security requirements of actual *RWprograms*. The axiomatic semantic, Hoare logic, is the method we choose to strengthen *propositions*. Because it is one of the strictest program verification methods in theorem proving technology and it is a method that can assist EVI to automatically verify *RWprogram* which will be explained later. Therefore, as rule Hoare-Inference, the following simplified axiomatic semantic definition is adopted in EVI. We take $m_{init}$ and $FRWprogram$ as preconditions and the $m_{final}$, which is satisfied expected properties, as postconditions. So the predicate $FInterpreter\ (m_{init}, FRWprogram) = m_{final}$ is the new stronger type / proposition needs to be verified. Obviously, only when the executing result of $FInterpreter\ (m_{init}, FRWprogram)$ is $m_{final}$, the proposition can be proved.

$$P\{m_{init}\}\ FInterpreter\ (m_{init}, FRWprogram)\ Q\{m_{final}\} \quad \text{(Hoare-Inference)}$$

As such, the stronger *propositions* are the relevant *properties* of *RWprograms*, and the respective proofs are the *verifications*. Accordingly, we obtain Corollary 2 and Corollary 3 below.

*properties as propositions as types* (Corollary 2)

*verifications as proofs* (Corollary 3)

Based on Corollaries 1, 2, and 3, we can deduce that the symbolic execution of an *RWprogram* in a formal system based on CHI is equivalent to constructing the respective proof terms of verification propositions, and vice versa. This is illustrated in Fig. 16. Thus, correspondences can be made still deeper yet to obtain a fourth principle below.

*verifications as execution of programs* (Principle 4)

Hereto, we have proven that EVI combines formal verification of higher-order theorem proving and symbolic execution technology.

## 5.2 Advantages of EVI

In this subsection, we summarize the advantages of the proposed EVI.

First of all, the proposed EVI, taking Hoare logic to strengthen *propositions*, deals the automatic verification problems of Hoare logic mentioned in Section 1. Specifically, if we want to use Hoare logic or similar style to automatically verify programs, we need to solve three problems: unifying the set of logic conditions, inferring intermediate logic conditions automatically, and proving loops. For solutions of the first and second problems, we take forward direction inference of Hoare logic as an example to illustrate them. Here is a very simple code segment and its verification using Hoare logic shown as below.

$$\text{"x := 1; y := x;"} \xrightarrow{prove} \text{"\{T\}x := 1; \{x = 1\}y := x; \{x = 1 \land y = 1\}"}$$

But if we add a new statement "z := x + y;", the verification process needs to be adjusted as following.

$$\text{"x := 1; y := 2; z := x + y;"} \xrightarrow{prove} \text{"\{T\}x := 1; \{x = 1\}y := 2; \{x = 1 \land y = 2 \land x + y = 3\}z := x + y; \{x = 1 \land y = 2 \land z = 3\}"}$$

This process can be abstracted as the process I, where we take $\bar{p}$ and $\bar{q}$ to represent the set of logic conditions of variables. It is obvious that if some logic statements $c_i$ are modified, the relevant sets of logic conditions $\bar{q}$ all need to be adjusted, including the number and definition of elements stored in $\bar{q}$. This chain reaction makes the manual verification workload involved become very heavy, and the discrete and non-uniform storage mode of $\bar{q}$ makes the automatic inferences become difficult and inefficient.

$$P\{\bar{p}\}c_0 \xrightarrow{infer} Q_0\{q_0^0 \dots q_j^0\}c_1 \xrightarrow{infer} Q_1\{q_0^1 \dots q_k^1\}c_2 \twoheadrightarrow c_n Q_n\{q_0^n \dots q_m^n\} \stackrel{?}{\leftrightarrow} Q\{\bar{q}_{final}\}, (j \leq k \leq m) \quad (I)$$

Actually, $\bar{p}$ and $\bar{q}$ is a type of indirect method to simulate the formal memory space. As rule II, with the help of GERM and EVI, $\bar{p}$ and $\bar{q}$ are encapsulated as the formal memory state $m$ that stores all variables information into their allocated memory blocks, and the intermediate memory states can be rebuilt by formal interpreter automatically instead of deriving manually.

Note that, as mentioned in subsection 5.1, we do not limit the range of formal semantics, because the distinctions among the operational semantics, denotational semantics and axiom semantics can be vague in the process of symbolic execution (verification). In $\mathcal{E}$, no matter what types of formal semantics chosen as the basis of the formal interpreter, they all can equivalently define the logic behaviors of IPL mathematically. The formal interpreter follows the semantics of IPL to generate logic abstract expressions for formal memory state, which can be used as the logic conditions in the Hoare style inferences.

$$P\{m_{init}\}c_0 \xrightarrow{FInterpreter(m_{init},c_0)} Q_0\{m_0\}c_1 \xrightarrow{FInterpreter(m_0,c_1)} Q_1\{m_1\}c_2 \twoheadrightarrow c_n Q_n\{m_n\} \stackrel{?}{\leftrightarrow} Q\{m_{final}\} \quad (II)$$

For proving loops problem, due to the combination of symbolic execution and higher-order theorem proving, we can use Bounded Model Checking (BMC) [26] and finding loop invariants simultaneously. At first, we employ BMC notion to set a limitation into the formal interpreter that the formal interpreter only can execute K times. In a general sense, if the execution of the *FRWprogram* can generate the corresponding final memory state using L times ($L \leq K$), it means the loops exist in the *FRWprogram* can be unfolded as a set of identical normal sequence statements directly in finite times which can be inferred by rule II. Then, if the execution of the *FRWprogram* invokes into a loop and cannot finish the loop after K times executing, we can set the loop statement as a break point, and separate the *FRWprogram* as two parts, head and tail. Next we need to find out the loop invariants and encapsulated them as invariant memory state $I\{m_i\}$, which plays the roles of the final memory state of the head part and the initial memory state of the tail part, as illustrated in process III.

$$P\{m_{init}\}c_0 \twoheadrightarrow c_i I\{m_i\} \, (head) \text{ and } c_i I\{m_i\} \twoheadrightarrow c_n Q\{m_{final}\} \, (tail) \quad (III)$$

And by employing composition rule of Hoare logic, we can get $P\{m_{init}\}c_0 \twoheadrightarrow c_n Q\{m_{final}\}$.

Hence, through GERM and EVI, as rule Hoare-Inference, although, the modification of logical statements will cause the changes of intermediate logic conditions, the definition of states does not need to be adjusted, and the new intermediate logic conditions stored in the each state can be inferred automatically. And therefore we can take Hoare logic to strengthen *propositions* and automatically verify *FRWprograms*.

Besides, EVI also can solve the consistence, reusability and automation problems.

- First, for consistence problems, according to Principle 4, the execution of *FRWprograms* written in IPL is isomorphic to their formal verifications. Therefore, obviously, *FRWprograms* play the role of the formal models of corresponding *RWprograms*. According to Relation 4, $RWprogram \equiv FRWprogram$, so the formal model is equivalent with target *RWprograms* without consistent problems. Besides, because $GPL \equiv IPL$, the modeling process is standardized as equivalently translating *RWprograms* into *FRWprograms* line by line mechanically without rebuilding, abstracting or any other steps which need to depend on the experience, knowledge, and proficiency of researchers. So it is also impossible to introduce consistence problems during constructing formal models. Furthermore, this mechanical translation process can be finished by specific translators automatically and reduce the workload caused by building formal models which also contributes to the automation problem.

- Second, for reusability problems, as rule Hoare-Inference, if we want to verify same theorems for different programs, instead of rebuilding the whole formal models, we only need to replace the $m_{init}$ and $FRWprogram$, vice versa, if we want to verify different theorems for same programs, we only need to replace the $m_{final}$. Besides the $m_{init}$, which have been verified, can directly be used in other verifications as $m_{init}$.

- Finally, for automation problems, it should be considered from two aspects. In theory, as mentioned above, the formal verification can be

finished automatically by symbolically executing *FRWprograms* in $\mathcal{E}$. In practice, the program verification process of all formal models based on EVI has been unified as the process of evaluating $FInterpreter\ (m_{init}, FRWprogram)$ and proving the equivalence between the result memory state and excepted final memory state. Thus the differences of program verification processes among different formal models have been reduced. Therefore it becomes possible to design *subtactics* based on the "*tactic*" mechanism provided by proof assistants which can finish different parts of the verification process, and combine them to become a large *tactic* which can finish the verification process fully automatically by employing the combination of *tactics*. In Table. 8, we illustrate the comparison of automation of building formal models, defining formal properties and verifying among VST, deep specifications framework and EVI.

Table 8. The comparison of automation about building formal models, defining properties and verifying among VST, deep specification framework and EVI.

|  | VST | Deep Specification framework | EVI |
|---|---|---|---|
| Modeling | Fully-automatic | Manually | Fully-Automatic |
| Defining Property | Manually | Manually | Manually |
| Verifying | Semi-automatic | Manually to Semi-automatic | Semi-automatic to Fully-automatic |

## 6. Advanced application of the GERM framework and EVI

In this section, we implement a toy FSPVM to emphasize the feasibility of our blueprint about the FSPVM based on GERM and EVI and concretely show the advantages mentioned above. For simulating the situation that four challenges have been overcome, we choose a simple imperative programming IMP [25], embodying a tiny core fragment of conventional mainstream languages such as C and Java, as the target GPL which has been used as an example language in many classical text books about programming language theory, and present the design of its IPL and the implementation of a corresponding formal interpreter in Coq based on the proposed formal memory model. Then we discuss their use for verifying security properties in Coq based on Principle 4.

### 6.1 Toy IPL

The abstract syntax of IMP is given in Table. 9, which only has basic arithmetic and Boolean datatypes. As a first step, we build a formal toy IPL based on IMP [25]. It is formally structured into types, values, expressions with variables, and statements, and its formal syntax and semantics are defined as strong type using generalized algebraic datatypes (GADTs) [2]. In this way, we give types of syntax constructors directly and it is impossible to construct ill-typed terms and stuck during evaluation in the type system of Coq. And the formal static and dynamic semantics are more easily to be defined and understood. A brief introduction to its formal abstract syntax is given in the following subsection.

Table 9. Abstract syntax of IMP.

| *Type:* | $\tau ::= nat\ \|\ bool$ |
|---|---|
| *Value:* | $a ::= \mathbb{N}$ |
|  | $b ::= true\ \|\ false$ |
| *Expression:* | $e ::= a\ \|\ b\ \|\ e_1 + e_2\ \|\ e_1 - e_2\ \|\ e_1 == e_2\ \|\ e_1\|\|e_2\ \|\ e_1 \&\& e_2\ \|\ id$ |
| *statement:* | $s ::= s_1\ ;;\ s_2\ \|\ if\ e\ then\ s_1\ else\ s_2\ \|\ e_1 = e_2\ \|\ skip\ \|\ throw$ |

#### 6.1.1 Formal abstract syntax

*Type*: This corresponds in IMP to the *type* signature incorporated in the data environment variable *Env* of the formal memory model, as described in Definition 2, and is employed in IMP to classify language values and expressions. Because IMP has only arithmetic and Boolean datatype values and variable expressions, the only types defined are *Tint*, *Tbool*, and *Tvid*. And these types are used as the type signatures to define the GADTs of $value_{IMP}$ and $expr_{IMP}$. Their definitions in Coq are shown in Fig. 17.

```
Inductive type : Type :=
  | Tnat
  | Tbool
  | Tvid : option address →type.
```

Figure 17. Type definitions of IMP in Coq

*Value*: This corresponds in the value of IMP to *value* in the formal memory model. In IMP, this term is formally defined as a GADT $value_{IMP}: (\forall t: type, val\ t)$, where it is parameterized by the type signature. According to Table. 9, the values of IMP include nature number and Boolean value, so $value_{IMP}$ constructed by constructor *Vnat* and *Vbool*. Its formal definition in Coq is shown in Fig. 18.

```
Inductive val : type → Type :=
  | Vnat : nat → val Tint
  | Vbool : bool → val Tbool.
```

Figure 18. Value definitions of IMP in Coq

*Expression*: Formal syntax of IMP expressions are also defined using GADTs as $expr_{IMP}: (\forall t_0\ t_1: type, expr_{t_0\ t_1})$. The $\tau_0$ refers to the expression current type and the $\tau_1$ refers to the type after evaluation. For instance, there is Boolean variable expression *e*, so the type of *e* is $expr_{Tvid\ (oa)\ Tbool}$. In this way, the formal syntax of expressions becomes more clear and abstract, and can keep the type safety of the IPL of IMP expressions strictly. In addition, it specifies and limits the semantics of each expression constructor employed for each of the three kinds of expressions given in Table. 9 representing constant, variable, and binary operation expressions, respectively. Their formal abstract syntax in Coq is defined in Fig. 19.

```
Inductive bop : type → type → Type :=
  | feqbOfNat  : bop_{Tnat Tbool}
  | fplusOfNat : bop_{Tnat Tnat}
  | fsubOfNat  : bop_{Tnat Tnat}
  | forbOfBool : bop_{Tbool Tbool}
  | fandbOfBool : bop_{Tbool Tbool}.
Inductive expr : type → type → Type:=
  | Econst : forall t, val t → expr_{t t}
  | Evar   : forall addr t, expr_{(Tvid addr) t}
  | Ebop   : forall t_0 t_1 t_2 t'_2, bop_{t_2 t'_2},
             expr_{t_0 t_2} → expr_{t_1 t_2} → expr_{t'_2 t'_2}.
```

Figure 19. Expression definitions of IMP in Coq

*Statement*: In IMP, this includes *conditional*, *assignment*, *sequence*, *skip*, and *throw* statements. Specially, *throw* is a type of statement that will halt the execution of the entire program, which is often defined in a JavaScript-like programming language such as Solidity. Besides, Benefiting from GADTs definition, formal statements of IMP are all well-formed. For example, informal syntax of IMP may occur

$$\text{if (``error'')}\ s_0\ s_1.$$

Although, these syntax errors will be found out during compiling, if we want to evaluate in higher-order theorem proving assistants, they may be stuck or figure out wrong result. But because of the type signatures, the type of condition has been limited as $\forall t_0: type, expr_{t_0\ Tbool}$ in formal abstract syntax tree. And it is impossible to construct ill-formed statements like the above one. Their formal abstract syntax in Coq is defined in Fig. 20.

```
Inductive statement : Type :=
  | If : forall t, expr_{t Tbool} → statement
         → statement → statement
  | Assignv : forall t t_0 t_1,
              expr_{t t_1} → expr_{t_0 t_1} → statement
  | Seq : statement → statement → statement
  | Snil : statement
  | Throw : statement.
```

Figure 20. Statement definitions of IMP in Coq

## 6.2 Toy formal interpreter

As a second step, we formalize the semantics of IMP and implement the formal interpreter in Coq, which can follow the formal semantics to interact with the GERM framework and evaluate a *FRWprogram* written in the IPL of IMP to its final memory state. This formal interpreter is very simple, and its correctness can be verified in Coq easily. Therefore, we do not evaluate the correctness of the interpreter here. A brief introduction to the formal semantics and respective formal interpreter functions are given in the following subsection.

### 6.2.1 Formal semantics

*Evaluation of value*: The semantics of value evaluation are transforming the values of IPL to the values of *memory*. According to *Value* syntax, the semantics can be defined easily as Fig. 21.

```
Definition val_to_value v_env b_infor t (v : val t) : option value :=
 match v with
   | Vnat n ⇒ Some (Nat (Some n) v_env b_infor)
   | Vbool b ⇒ Some (Bool (Some b) v_env b_infor)
 end.
```

Figure 21. Implementation of evaluation of value in Coq

The formal semantics of expressions are defined separately as *expression for left value* and *expression for right value* [40], which is used to distinguish the typical mode of value or expression evaluation on the left and right hand side of an assignment statement.

*Evaluation of expression for left value*: This kind of expression semantics is used to evaluate the expressions at the left side of assignment statements. In IMP, only variable expressions can serve as the left value of assignment statements, thus the evaluation process is very simple, where, if the current expression is a variable expression, then the indexed label address is returned; else, an error result is returned. The implementation in Coq is shown in Fig. 22.

```
Fixpoint expr_l t_0 t_1 (e : expr_{t_0 t_1}) : option address :=
 match e with
   | Evar oaddr t_1 ⇒ oaddr
   | _ ⇒ None
 end.
```

Figure 22. Implementation of evaluation of expression left value in Coq

*Evaluation of expression for right value*: For an expression evaluated as a right value, all kinds of IMP expressions can serve as the right value of assignment statements. A constant expression will be evaluated according to the semantics of *Evaluation of value*, where the variable expression will access the respective memory block using $read_{chck}$, and a binary operation expression will be evaluated by assistant functions extended in Coq according to the value of the *bop* type given in Fig. 19. The implementation in Coq is shown in Fig. 23.

```
Fixpoint expr_r t_0 t_1 m v_env b_infor (e : expr_{t_0 t_1}) : option value :=
 match e with
   | Econst v ⇒ (val_to_value m v_env v)
   | Evar oaddr t_1 ⇒
     match oaddr with
       | None ⇒ None | Some addr ⇒ (read_{chck} m v_env b_infor addr)
     end
   | Ebop b e_0 e_1 ⇒
     match b with
       | feqbOfNat ⇒ eqb_val (expr_r m v_env b_infor e_0) (expr_r m v_env b_infor e_1)
       | fplusOfNat ⇒ plus_val (expr_r m v_env b_infor e_0) (expr_r m v_env b_infor e_1)
       | fsubOfNat ⇒ sub_val (expr_r m v_env b_infor e_0) (expr_r m v_env b_infor e_1)
       | forbOfBool ⇒ orb_val (expr_r m v_env b_infor e_0) (expr_r m v_env b_infor e_1)
       | fandbOfBool ⇒ andb_val (expr_r m v_env b_infor e_0) (expr_r m v_env b_infor e_1)
     end end.
```

Figure 23. Implementation of evaluation of expression right value in Coq

*Evaluation of statement*: This is the entry point of the formal interpreter. It takes as parameters a memory state $m_{init}$, current environment information $v_{env}$, current block information $b_{infor}$, and the *FRWprogram* to be executed. Then, it evaluates *program* into a final memory state $m_{final}$. The implementation in Coq is shown in Fig. 24.

```
Fixpoint test K m v_env b_infor (stt : statement) : memory :=
match K with
 | O ⇒ m | S K' ⇒
 match m.(m_throw) with
   | true ⇒ init_m | false ⇒
    match stt with
      | Snil ⇒ m | Throw ⇒ (write_dir m _0xthrow true)
      | Seq s_0 s_1 ⇒ let m' := test K' m v_env b_infor s_0 in
                            test K' m' v_env b_infor s_1
   | If e s_0 s_1 ⇒
      match expr_r v_env b_infor e with
        | None ⇒ None | Some v ⇒
         match v with
          | Bool b _ _ ⇒
           match b with
             | Some true ⇒ test K' m v_env b_infor s_0
             | Some false ⇒ test K' m v_env b_infor s_1
             | None ⇒ m
           end
          | _ ⇒ m
         end end
   | Assignv e_0 e_1 ⇒
      match expr_r v_env b_infor e_1 with
       | None ⇒ m
       | Some v ⇒ match expr_l v_env b_infor e_0 with
                   | None ⇒ m | Some addr ⇒ write_chck m v_env b_infor addr v
                  end
end end end end.
```

Figure 24. Implementation of evaluation of statement in Coq

**6.3 Case study: automation and verification**

    Based on above preparation work, we have constructed a toy FSPVM for IMP. We can use it to verify programs written in IMP. Here we give a simple example code in the left of Fig. 26. The requirement of it is that, if *Pledge* is zero or *complete* or *refunded* is true, then the *throw* statement is executed; else, the variable *refnd* is assigned as true. In practice, the process of verification based on EVI involves three steps: 1) defining a formal model; 2) defining properties (including setting initial and final memory space); 3) verifying (executing). For step 1, the process of modeling can be finished completely automatically by assistant tools such translator. And the result is given in the right of Fig. 26. Step 2 needs to be completed manually according to the requirements of code. Automating this process is a difficult problem that remains unsolved in nearly all existing theorem proving and model checking assistant applications. Specifically, first, we need to initialize the variables in conditional statement in $m_{init}$, and set the postcondition that if the condition is true, then the value of *refnd* stored in $m_{final}$ should be *true*, else $m_{init} \equiv m_{final}$. Finally, step 3 is almost fully automatically that is illustrated by the example given in Fig. 25. Here, the cases of abstract values are classified by the *destruct* procedure, and then the *step* procedure, which is designed through the "*tactic*" mechanism of Coq, is applied to complete the verification automatically. The verification process based on EVI is conducted as an evaluation by the interpreter, and the process is fixed. Therefore, the *step* procedure can contain all of the inference rules of the evaluation process. As such, the *step* procedure is like a switch launching the interpreter to symbolically execute any formal *FRWprogram*, and then generating the proof terms of the respective properties automatically. The only one manual substep in step 3 is classifying the cases of abstract values by the *destruct* procedure. However, we should note that this substep has a fixed rule that only abstract values defined in $m_{init}$ are needed to be processed. So this substep is able to be solved automatically, and we are going to deal it. Moreover, we can set a break point manually to observe logic invariants stored in intermediate memory states, as shown in Fig. 27.

```
Theorem _pledge_correct : forall b₁ b₂ n m₀ m₁ m₂ v_env b_infor,
m₀ = write_dir m_init (Bool (Some b₁) v_env b_infor) _complete →
m₁ = write_dir m₀ (Bool (Some b₂) v_env b_infor) _refunded →
m₂ = write_dir m₁ (Nat (Some n) v_env b_infor) _Pledge →
((z = 0 ∨ b₁ = true ∨ b₂ = true)
∧ test K m₂ v_env b_infor _pledge_test = m_init) ∨
((z ≠ 0 ∧ b₁ = false ∧ b₂ = false)
∧ read_dir (test K m₂ v_env b_infor _pledge_test) _refnd = (Bool (Some true) v_env b_infor))
Proof.
 destruct v_env, b_infor; intros; initmem.
 rewrite H1.
 destruct n, b₁, b₂; step.
Qed.
```

Figure 25. Example of the verification step (step 3)

```
if (Pledge == 0 ||           Definition _pledge_test :=
   complete ||                  Seq (If ((Evar (Some _Pledge) Tnat)
   refunded ) {                       (==) (Econst (Vnat 0))
    throw; }                           (||) (Evar (Some _complete) Tbool)
refnd = true;                          (||) (Evar (Some _refunded) Tbool))
                                      (Throw) (Snil))
                                (Seq (Assignv (Evar (Some _refnd) Tbool)
                                              (Econst (Vbool true)))
                                Snil).
```

Figure 26. Simple *RWprogram* code segment (right) and corresponding formal program (left)

According to above simple example, we illustrate feasibility to build a FSPVM and verify programs in practice based on GERM and EVI. Actually, we have employed GERM and EVI into our ongoing project for Ethereum smart contract verification and the Fig. 28 shows an example of it. Although it is much more complex than the above example given in Fig. 17 to 27, it is clear that they have identical verification process, which also ensures the feasibility of the FPSVM blueprint. And these relevant works about our project will be introduced in our other papers when they are completed.

Figure 27. Formal memory state during verify

Figure 28. An verification example of the ongoing project about building a FSPVM for Ethereum smart contracts using GERM and EVI

**7. Conclusions and future work**

**7.1 Conclusion**

In this paper, we developed a general, extensible, and reusable formal memory (GERM) framework based on the calculus of inductive constructions, and implemented and verified the framework in Coq. This independent and customizable framework is employed to simulate the structure and operations of physical memory hardware, and provides a basis for users to easily construct formal models of programs written in any high-level language for program verification. We also presented an extension of Curry-Howard isomorphism, denoted as execution-verification isomorphism (EVI), which combines symbolic execution and theorem proving for solving the problems of automation of verification in higher-order logic theorem proving assistant tools, and. Finally, we define a toy FSPVM, including a toy language and a respective formal interpreter in Coq based on the GERM framework and EVI notion, and verify a simple code segment in order to demonstrate the feasibility and advantages of the blueprint of our expected FSPVM.

**7.2 Future work**

We are presently pursuing the formalization of higher-level smart contract development languages, including Serpent [27] and Solidity [28]. We will then develop a formal verified interpreter for these languages based on the GERM framework. Finally, we will build a general formal verification toolchain for blockchain smart contracts based on EVI with the goal of developing automatic smart contract verification.

**Acknowledgements**

The authors thank Marisa, Yan Fang and Yunzhuang Cai for their kind assistance.